\documentclass[showkeys,superscriptaddress,floatfix,aps,10pt,prd]{revtex4-2}
\usepackage{graphicx,epstopdf}
\usepackage[caption=false]{subfig}
\usepackage{appendix}
\usepackage[T1]{fontenc}
\usepackage{lmodern}
\usepackage[dvipsnames,x11names]{xcolor}
\usepackage[colorlinks=true,linkcolor=ForestGreen,citecolor=ForestGreen,urlcolor=NavyBlue]{hyperref}
\usepackage[sort&compress]{natbib}
\usepackage{lipsum}
\usepackage{morefloats}
\usepackage[pdf]{pstricks}
\usepackage{amsmath}
\usepackage{amssymb}
\usepackage{amsfonts}
\usepackage{rotating}
\usepackage{cancel}
\usepackage{mathtools}
\usepackage{bbm}
\usepackage{dsfont}
\usepackage{bbold}
\usepackage{multirow}
\usepackage{ulem}
\usepackage{physics}
\usepackage{orcidlink}
\usepackage{colortbl}

\def\xslash{x\!\!\!\slash }

\def\vel{\left|}
\def\ver{\right|}

\begin{document}

\title{Elucidating the nature of axial-vector charm-antibottom tetraquark states}

\author{Ula\c{s}~\"{O}zdem\orcidlink{0000-0002-1907-2894}}%
\email[]{ulasozdem@aydin.edu.tr }
\affiliation{ Health Services Vocational School of Higher Education, Istanbul Aydin University, Sefakoy-Kucukcekmece, 34295 Istanbul, T\"{u}rkiye}

 
\begin{abstract}
Investigating the electromagnetic characteristics of unconventional states may offer new insights into their internal structures. In particular, the magnetic moment attributes may serve as a crucial physical observable for differentiating exotic states with disparate configurations or spin-parity quantum numbers. As a promising avenue for research, encompassing both opportunities and challenges, an in-depth examination of the electromagnetic properties of exotic states is crucial for advancing our understanding of unconventional states. Motivated by this, in this study, the magnetic moments of $ \rm{I(J^{P})} = 1(1^{+})$ $Z_{\bar b c}$  tetraquark states are analyzed in the framework of QCD light-cone sum rules by considering the diquark-antidiquark approximation, designated as type $3_c \otimes \bar 3_c$.  Although the tetraquark states examined in this study have nearly identical masses, their magnetic moment results exhibit noticeable discrepancies. This may facilitate the differentiation between quantum numbers associated with states with identical quark content. The results show that heavy quarks overcoming light quarks can determine both the sign and the magnitude of the magnetic moments of these tetraquark states.  The numerical results obtained in this study suggest that the magnetic moments of $Z_{\bar b c}$ tetraquark states may reveal aspects of their underlying structure, which could distinguish between their spin-parity quantum numbers and their internal structure. The results obtained regarding the magnetic moments of the $Z_{\bar b c}$  tetraquark states may be checked within the context of different phenomenological approaches.
\end{abstract}

\maketitle

\section{motivation}\label{motivation}

Although the existence of hadrons with more sophisticated configurations than those comprising $q\bar q$ and $qqq$ has been known for several decades, the experimental confirmation of the presence of the exotic state labeled X(3872) was accomplished by the Belle Collaboration in 2003~\cite{Belle:2003nnu}.  Subsequently, various experimental collaborations have identified a considerable number of exotic states. With each new observation, the family of such exotic states continues to diversify, representing a dynamically evolving and active field within hadron physics, encompassing experimental and theoretical approaches. Various theoretical explanations have been put forth to elucidate the true nature of these states. These explanations have been proposed in conjunction with several different theoretical constructs, including more traditional hadrons, loosely bound molecular states, compact pentaquarks or tetraquarks, hybrids, glueballs, kinematic effects, and other related concepts. However, despite comprehensive investigations, both theoretical and experimental, the fundamental questions concerning the nature, quantum numbers, and decay properties of these exotic states remain unresolved. 
A list of the most recent advances in the domain of exotic states can be found in Refs.~\cite{Esposito:2014rxa, Esposito:2016noz,Olsen:2017bmm, Lebed:2016hpi,Nielsen:2009uh, Brambilla:2019esw,Agaev:2020zad, Chen:2016qju,Ali:2017jda, Guo:2017jvc,Liu:2019zoy, Yang:2020atz,Dong:2021juy, Dong:2021bvy, Meng:2022ozq, Chen:2022asf}. 

Most observed tetraquark states are classified as hidden-charm or hidden-bottom tetraquark states, encompassing the $ c \bar c $ or $b \bar b$ pair. Nevertheless, the fundamental principles of QCD do not preclude the theoretical possibility of open-flavour tetraquark states. The $[qc] [\bar q \bar b]$ with $q$ $=$ $u$, $d$ and $s$  tetraquark states represent a distinct category of exotic state. A considerable amount of theoretical information on the expected properties of these tetraquark states is available in the literature since the original studies began more than two decades ago~\cite{Zhang:2009vs, Zhang:2009em, Sun:2012sy, Albuquerque:2012rq, Chen:2013aba, Agaev:2016dsg, Agaev:2017uky, Wang:2020jgb, Wang:2019xzt, Wu:2018xdi, Ortega:2020uvc}. While these specific tetraquark configurations have not yet been observed experimentally, they represent a theoretically compelling subject in hadron spectroscopy. The $[qc][\bar{q}\bar{b}]$ system is particularly noteworthy because, unlike hidden-flavor tetraquarks (e.g., $c\bar{c}q\bar{q}$ or $b\bar{b}q\bar{q}$ ), it possesses inherent stability and is expected to exhibit narrow decay widths. This distinctive feature arises from the flavor-asymmetric quark content, which forbids annihilation into gluons—a dominant decay channel for many conventional quarkonia and hidden-flavor exotic states \cite{Ortega:2020uvc}. These properties render them a valuable resource for the study of heavy-quark dynamics and a deeper understanding of the dynamics of QCD. To gain a deeper understanding of the internal structures of these states, it is also crucial to investigate the decay channels, including strong, radiative, and electromagnetic, in conjunction with their spectroscopic parameters.

The study of the electromagnetic properties of tetraquark states provides a crucial window into their internal quark-gluon structure and serves as a powerful probe of non-perturbative QCD dynamics. Among these properties, the magnetic dipole moment stands out as a particularly sensitive observable, as it directly reflects the spatial distribution of charges and spins within the hadron. This work aims to determine the magnetic moments of the axial-vector charm-antibottom tetraquark states, denoted as $[qc][\bar q^\prime \bar b]$ and $[qc][\bar q \bar b]$ (for short $Z_{\bar b c}$) using the framework of QCD light-cone sum rules (LCSR). We construct these states in a compact diquark-antidiquark configuration with spin-parity quantum numbers $\mathrm{J^P} = 1^{+}$. Recent experimental discoveries have revealed a rich spectrum of exotic hadrons, with many states exhibiting properties consistent with loosely bound molecular configurations of conventional mesons and baryons. While this molecular picture is phenomenologically successful for numerous candidates, it is crucial to recognize that QCD fundamentally permits another distinct class of exotic states: compact tetraquarks, where all four valence quarks are confined within a single, small-volume color-singlet object. The central challenge in contemporary hadron spectroscopy lies not in choosing between these two paradigms, but in developing tools to distinguish them unambiguously. In this context, electromagnetic properties, particularly the magnetic dipole moment, emerge as exceptionally sensitive and discriminating probes. Unlike the mass, which can be similar for molecular and compact configurations due to intricate binding dynamics, the magnetic moment is directly governed by the intrinsic spatial distribution of charges and spins within the hadron. A compact diquark-antidiquark system, characterized by strongly correlated quark pairs in a small volume, is expected to produce a magnetic response fundamentally different from that of a loosely bound hadronic molecule, where the constituent mesons largely preserve their individual electromagnetic identities. Therefore, precise theoretical predictions for the magnetic moments of specific tetraquark configurations establish critical theoretical benchmarks against which future experimental measurements can be compared. Consequently, this work is dedicated to providing first-principles predictions for the magnetic moments of the axial-vector $Z_{\bar b c}$tetraquark states within the compact diquark-antidiquark picture, using the method of LCSR. Our calculations yield concrete numerical predictions that serve as a reference for the compact scenario. Should future high-precision experiments at facilities like LHCb, Belle II, or a future electron-positron super factory succeed in measuring the electromagnetic properties of a charged $Z_{\bar b c}$-like resonance, a direct comparison with our results could provide compelling evidence for or against its interpretation as a compact tetraquark. This strategy of utilizing electromagnetic observables to disentangle competing structural models represents a vital and necessary direction for advancing our understanding of exotic hadron structure. This research extends the systematic investigation of electromagnetic properties into the open-charm-bottom tetraquark sector. It builds upon a substantial body of prior work examining the electromagnetic characteristics of both hidden- and doubly-heavy tetraquark states~\cite{Ozdem:2024txt,Ozdem:2024dbq, Ozdem:2024lpk, Mutuk:2023oyz, Wang:2023bek, Ozdem:2023rkx, Ozdem:2023frj, Lei:2023ttd, Zhang:2021yul, Azizi:2023gzv, Ozdem:2022eds, Ozdem:2022kck, Xu:2020qtg, Wang:2017dce, Ozdem:2022yhi, Wang:2023vtx, Ozdem:2021hmk, Azizi:2021aib, Ozdem:2021hka, Xu:2020evn, Ozdem:2021yvo, Ozdem:2017exj, Ozdem:2017jqh, Ozdem:2024rrg, Mutuk:2024vzv} and pentaquark states~\cite{Wang:2016dzu, Ortiz-Pacheco:2018ccl, Xu:2020flp, Li:2021ryu, Ozdem:2023htj, Wang:2023iox, Ozdem:2022kei, Gao:2021hmv, Ozdem:2024rch, Guo:2023fih, Ozdem:2022iqk, Wang:2022nqs, Wang:2022tib, Ozdem:2024jty, Li:2024wxr, Li:2024jlq, Ozdem:2024yel, Ozdem:2024rqx, Mutuk:2024ltc, Mutuk:2024jxf, Zhu:2025abk, Ozdem:2025ncd, Ozdem:2025ion, Mutuk:2024ach, Ozdem:2025jda}. It is evident that further research is required to gain a deeper understanding of the electromagnetic characteristics of states with varying structural compositions.

  The work is arranged as follows: Section~\ref{formalism} is concerned with deriving the LCSR for the magnetic moment of $Z_{\bar b c}$ states with quantum numbers $ \rm{J^{P}}= 1^{+}$.  Section~\ref{numerical} is dedicated to the numerical examination of the sum rules derived for these states. In Section \ref{sec:experimental}, possible scenarios for the experimental measurement of the obtained magnetic and quadrupole moments are presented. The final section of this study is devoted to the presentation of our concluding remarks.

 \begin{widetext}
 
\section{Construction of the LCSR}\label{formalism}
This section of the paper presents the derivation of LCSR for $Z_{\bar b c}$ states. To obtain the relevant sum rules, the following correlation function is considered. 
\begin{equation}
 \label{edmn01}
\Pi _{\mu \nu }(p,q)=i\int d^{4}x\,e^{ip\cdot x}\langle 0|\mathcal{T}\{J^i_{\mu}(x)
J_{\nu }^{i \dagger }(0)\}|0\rangle_{F}, 
\end{equation}
where $F$ denotes the external electromagnetic field, while $J^i_{\mu(\nu)}(x)$ denotes the interpolating currents, which correspond to the $Z_{ \bar b c}$ tetraquark states. 


Under the prescriptions of the LCSR, the analysis processes can be written as follows:
\begin{itemize}
 \item The correlation function is derived in terms of hadronic parameters, including mass, residue, form factors, and so forth, which is referred to as the "hadronic representation".
 
 \item The correlation function is also derived in terms of parameters associated with the QCD parameters, including quark-gluon degrees of freedom and distribution amplitudes, which is referred to as the "QCD representation".
 
 \item In the final step, the aforementioned representations are equalized with the help of the quark-hadron duality assumption. To remove any unwanted contributions from the analyses, a double Borel transformation and continuum subtractions are performed to obtain sum rules for the physical parameter to be calculated.
\end{itemize}

In accordance with the procedure described above, our analysis starts by examining the hadronic formulation of the relevant tetraquark states.
\subsection{Hadronic representation of the correlation function}

By the prescription outlined above, we may now commence the magnetic moment analyses of $Z_{\bar b c}$ tetraquark states.  To obtain the hadronic representation of the correlation function,  we plug into the correlation function a complete set of intermediate $Z_{\bar b c}$ tetraquark states with the same quantum numbers as the interpolating currents and then taking the integral over x, we get following results:
\begin{align}
\label{edmn04}
\Pi_{\mu\nu}^{Had} (p,q) &= {\frac{\langle 0 \mid J_\mu (x) \mid
Z_{\bar b c}(p,\varepsilon^i) \rangle}{p^2 - m_{Z_{\bar b c}}^2}} \langle Z_{\bar b c} (p, \varepsilon^i) \mid Z_{\bar b c} (p+q, \varepsilon^f) \rangle_F 
\frac{\langle Z_{\bar b c} (p+q, \varepsilon^f) \mid J_{\nu }^{\dagger } (0) \mid 0 \rangle}{(p+q)^2 - m_{Z_{\bar b c}}^2} \nonumber\\
&+ \mbox{higher states and continuum}.
\end{align}

In Eq. (\ref{edmn04}), there are matrix elements whose explicit forms are required, which are written as follows~\cite{Brodsky:1992px}:
\begin{align}
\label{edmn05}
\langle 0 \mid J_\mu (x) \mid Z_{\bar b c} (p, \varepsilon^i) \rangle &=  \lambda_{Z_{\bar b c}} \varepsilon_\mu^i\,,\\
\langle Z_{\bar b c} (p+q, \varepsilon^{f}) \mid J_{\nu }^{\dagger } (0) \mid 0 \rangle &= \lambda_{Z_{\bar b c}} \varepsilon_\nu^{* f}\,,\\
\langle Z_{\bar b c}(p,\varepsilon^i) \mid  Z_{\bar b c} (p+q,\varepsilon^{f})\rangle_F &= - \varepsilon^\gamma (\varepsilon^{i})^\mu (\varepsilon^{f})^\nu
\Big[ G_1(Q^2)~ (2p+q)_\gamma ~g_{\mu\nu}  
+ G_2(Q^2)~ ( g_{\gamma\nu}~ q_\mu -  g_{\gamma\mu}~ q_\nu)
\nonumber\\ 
&
- \frac{1}{2 m_{Z_{\bar b c}}^2} G_3(Q^2)~ (2p+q)_\gamma
q_\mu q_\nu  \Big],\label{edmn06}
\end{align}
where $\lambda_{Z_{\bar b c}}$ and $ \varepsilon_\mu^i (\varepsilon_\nu^{*f}) $   are the residue, and the polarization vector of the initial and final $Z_{\bar b c}$ tetraquark states, respectively; $\varepsilon^\gamma$ is the polarization vector of the photon, and  $G_i(Q^2)$'s are Lorentz invariant form factors of the corresponding radiative transition with  $Q^2=-q^2$.

Upon substituting the Eqs.~(\ref{edmn05})-(\ref{edmn06}) into Eq.~(\ref{edmn04}), the hadronic representation of the correlation function of the $Z_{\bar b c}$ tetraquark states takes the following form:
%
%
\begin{align}
\label{edmn09}
 \Pi_{\mu\nu}^{Had}(p,q) &=  \frac{\varepsilon_\rho \, \lambda_{Z_{\bar b c}}^2}{ [m_{Z_{\bar b c}}^2 - (p+q)^2][m_{Z_{\bar b c}}^2 - p^2]}
 \Bigg\{G_1(Q^2)(2p+q)_\rho\Bigg[g_{\mu\nu}-\frac{p_\mu p_\nu}{m_{Z_{\bar b c}}^2}
 -\frac{(p+q)_\mu (p+q)_\nu}{m_{Z_{\bar b c}}^2}+\frac{(p+q)_\mu p_\nu}{2m_{Z_{\bar b c}}^4}\nonumber\\
 & \times (Q^2+2m_{Z_{\bar b c}}^2)
 \Bigg]
 + G_2 (Q^2) \Bigg[q_\mu g_{\rho\nu}  
 - q_\nu g_{\rho\mu}-
\frac{p_\nu}{m_{Z_{\bar b c}}^2}  \big(q_\mu p_\rho - \frac{1}{2}
Q^2 g_{\mu\rho}\big) 
+
\frac{(p+q)_\mu}{m_{Z_{\bar b c}}^2}  \big(q_\nu (p+q)_\rho+ \frac{1}{2}
Q^2 g_{\nu\rho}\big) 
\nonumber\\
&-  
\frac{(p+q)_\mu p_\nu p_\rho}{m_{Z_{\bar b c}}^4} \, Q^2
\Bigg]
-\frac{G_3(Q^2)}{m_{Z_{\bar b c}}^2}(2p+q)_\rho \Bigg[
q_\mu q_\nu -\frac{p_\mu q_\nu}{2 m_{Z_{\bar b c}}^2} Q^2 
+\frac{(p+q)_\mu q_\nu}{2 m_{Z_{\bar b c}}^2} Q^2
-\frac{(p+q)_\mu q_\nu}{4 m_{Z_{\bar b c}}^4} Q^4\Bigg]
\Bigg\}\,.
\end{align}

The magnetic form factor, designated as $F_M(Q^2)$, can be derived under the established methodology for deriving the aforementioned form factors, $G_i(Q^2)$, in the following way:
\begin{align}
\label{edmn07}
&F_M(Q^2) = G_2(Q^2)\,.\nonumber \\
\end{align}

In the static limit ($Q^2=0$), where the photon is regarded as a real particle, the magnetic moment ($\mu_{Z_{\bar b c}}$) can be characterized as follows:
\begin{align}
\label{edmn08}
 \mu_{Z_{\bar b c}}  &= F_M(0) \,\Big(\frac{e}{2 m_{Z_{\bar b c}}} \Big)=F_M(0) \,\Big(\frac{m_N}{ m_{Z_{\bar b c}}}\Big)\mu_N ,  
\end{align}
where $m_N$ is mass of the nucleon and, $\mu_N$ stands for the nuclear magneton. 

The final equation has been obtained, thus enabling the hadronic representation of the analysis to be derived. The second step of the aforementioned prescription may now be initiated, namely the derivation of the correlation function in terms of QCD parameters.

 \subsection{QCD Representation of the correlation function}

 To compute the magnetic moments of the axial-vector tetraquark states via LCSR, one must first construct appropriate interpolating currents that couple efficiently to the desired physical states. These currents should properly reflect the internal diquark-antidiquark structure and the $J^P = 1^+$ quantum numbers of the states under study. In the present work, we adopt four independent interpolating currents that are built from a scalar diquark ($S$) and an axial-vector antidiquark ($A$), or vice‑versa. Such combinations are known to be the most favorable configurations for tetraquark states in the QCD sum rule approach~\cite{Wang:2010sh, Kleiv:2013dta}. The explicit forms of the currents are given by~\cite{Wang:2019xzt}:
\begin{align}
\label{curr1}
J_{\mu }^{1}(x) &= \frac{\varepsilon^{abc} \varepsilon^{ade}}{\sqrt{2}} \Bigg\{ \big[u^{b^T} (x) C \gamma_5 c^c (x)\big]\big[\bar d^d (x) \gamma_\mu C \bar b^{e^T} (x) \big] - \big[u^{b^T} (x) C \gamma_\mu c^c (x)\big]\big[\bar d^d (x) \gamma_5 C \bar b^{e^T} (x) \big]\Bigg\},\\
J_{\mu }^{2}(x) &= \frac{\varepsilon^{abc} \varepsilon^{ade}}{\sqrt{2}} \Bigg\{ \big[u^{b^T} (x) C \gamma_5 c^c (x)\big]\big[\bar d^d (x) \gamma_\mu C \bar b^{e^T} (x) \big] + \big[u^{b^T} (x) C \gamma_\mu c^c (x)\big]\big[\bar d^d (x) \gamma_5 C \bar b^{e^T} (x) \big]\Bigg\},\\
J_{\mu }^{3}(x) &= \frac{\varepsilon^{abc} \varepsilon^{ade}}{2} \Bigg\{  \big[u^{b^T} (x) C \gamma_5 c^c (x)\big]\big[\bar u^d (x) \gamma_\mu C \bar b^{e^T} (x) \big] - \big[d^{b^T} (x) C \gamma_5 c^c (x)\big]\big[\bar d^d (x) \gamma_\mu C \bar b^{e^T} (x) \big]  \nonumber\\
&- \big[u^{b^T} (x) C \gamma_\mu c^c (x)\big]\big[\bar u^d (x) \gamma_5 C \bar b^{e^T} (x) \big] + \big[d^{b^T} (x) C \gamma_\mu c^c (x)\big]\big[\bar d^d (x) \gamma_5 C \bar b^{e^T} (x) \big]\Bigg\},\\
J_{\mu }^{4}(x) &= \frac{\varepsilon^{abc} \varepsilon^{ade}}{2} \Bigg\{ \big[u^{b^T} (x) C \gamma_5 c^c (x)\big]\big[\bar u^d (x) \gamma_\mu C \bar b^{e^T} (x) \big] - \big[d^{b^T} (x) C \gamma_5 c^c (x)\big]\big[\bar d^d (x) \gamma_\mu C \bar b^{e^T} (x) \big]  \nonumber\\
&+ \big[u^{b^T} (x) C \gamma_\mu c^c (x)\big]\big[\bar u^d (x) \gamma_5 C \bar b^{e^T} (x) \big] - \big[d^{b^T} (x) C \gamma_\mu c^c (x)\big]\big[\bar d^d (x) \gamma_5 C \bar b^{e^T} (x) \big]\Bigg\},
\label{curr4}
\end{align}
where $a$, $b$, $c$, $d$ and $e$ are color indices, and $C$ denotes the charge conjugation operator.  

With these currents at hand, we proceed to derive the QCD representation of the correlation function in Eq.~(\ref{edmn01}). Substituting the interpolating currents for the initial and final tetraquark states and applying Wick's theorem yields the complete set of quark-field contractions. For illustration, the resulting expressions corresponding to the currents $J_\mu^1(x)$ and $J_\mu^2(x)$ are presented as: 
\begin{align}
\Pi _{\mu \nu }^{\mathrm{QCD},\,J_\mu^1}(p,q)&=\mathbb C \int d^{4}xe^{ip\cdot x} 
 \langle 0 \mid \bigg\{ 
 \mathrm{Tr}
\Big[ \gamma _{\mu }{S}_{b}^{e^{\prime }e}(-x)\gamma _{\nu} \widetilde S_{d}^{d^{\prime }d}(-x)\Big]    
\mathrm{Tr}\Big[ \gamma _{5} {S}_{c}^{cc^{\prime}}(x)\gamma _{5} \widetilde S_{u}^{bb^{\prime }}(x)\Big] 
\nonumber\\
&-\mathrm{Tr}
\Big[ \gamma _{\mu }{S}_{b}^{e^{\prime }e}(-x)\gamma _{5} \widetilde S_{d}^{d^{\prime }d}(-x)\Big]    
\mathrm{Tr}\Big[ \gamma _{5} {S}_{c}^{cc^{\prime}}(x)\gamma _{\nu} \widetilde S_{u}^{bb^{\prime }}(x)\Big]  
\nonumber \\
&-\mathrm{Tr}
\Big[ \gamma _{5 }{S}_{b}^{e^{\prime }e}(-x)\gamma _{\nu} \widetilde S_{d}^{d^{\prime }d}(-x)\Big]    
\mathrm{Tr}\Big[ \gamma _{\mu} {S}_{c}^{cc^{\prime}}(x)\gamma _{5} \widetilde S_{u}^{bb^{\prime }}(x)\Big]  
\notag \\
& +\mathrm{Tr}
\Big[ \gamma _{5}{S}_{b}^{e^{\prime }e}(-x)\gamma _{5} \widetilde S_{d}^{d^{\prime }d}(-x)\Big]    
\mathrm{Tr}\Big[ \gamma _{\mu } {S}_{c}^{cc^{\prime}}(x)\gamma _{\nu } \widetilde S_{u}^{bb^{\prime }}(x)\Big]  \bigg\} \mid 0 \rangle_{F} ,  \label{eq:QCDSide}
\\
\Pi _{\mu \nu }^{\mathrm{QCD},\,J_\mu^2}(p,q)&=\mathbb C \int d^{4}xe^{ip\cdot x} 
\langle 0 \mid \bigg\{ 
 \mathrm{Tr}
\Big[ \gamma _{\mu }{S}_{b}^{e^{\prime }e}(-x)\gamma _{\nu} \widetilde S_{d}^{d^{\prime }d}(-x)\Big]    
\mathrm{Tr}\Big[ \gamma _{5} {S}_{c}^{cc^{\prime}}(x)\gamma _{5} \widetilde S_{u}^{bb^{\prime }}(x)\Big] 
\nonumber\\
&+\mathrm{Tr}
\Big[ \gamma _{\mu }{S}_{b}^{e^{\prime }e}(-x)\gamma _{5} \widetilde S_{d}^{d^{\prime }d}(-x)\Big]    
\mathrm{Tr}\Big[ \gamma _{5} {S}_{c}^{cc^{\prime}}(x)\gamma _{\nu} \widetilde S_{u}^{bb^{\prime }}(x)\Big]  
\nonumber \\
&+\mathrm{Tr}
\Big[ \gamma _{5 }{S}_{b}^{e^{\prime }e}(-x)\gamma _{\nu} \widetilde S_{d}^{d^{\prime }d}(-x)\Big]    
\mathrm{Tr}\Big[ \gamma _{\mu} {S}_{c}^{cc^{\prime}}(x)\gamma _{5} \widetilde S_{u}^{bb^{\prime }}(x)\Big]  
\notag \\
& +\mathrm{Tr}
\Big[ \gamma _{5}{S}_{b}^{e^{\prime }e}(-x)\gamma _{5} \widetilde S_{d}^{d^{\prime }d}(-x)\Big]    
\mathrm{Tr}\Big[ \gamma _{\mu } {S}_{c}^{cc^{\prime}}(x)\gamma _{\nu } \widetilde S_{u}^{bb^{\prime }}(x)\Big]  \bigg\} \mid 0 \rangle_{F} ,  \label{eq:QCDSide2}
\end{align}%
where  $\mathbb C=  \frac{i}{2}\varepsilon^{abc}\varepsilon^{a^{\prime}b^{\prime}c^{\prime}}\varepsilon^{ade}
\varepsilon^{a^{\prime}d^{\prime}e^{\prime}}$ and $\widetilde{S}_{Q(q)}^{ij}(x)=CS_{Q(q)}^{ij\rm{T}}(x)C$. The relevant light ($S_{q}(x)$) and heavy ($S_{Q}(x)$) quark propagators in the presence of the external background field are written as~\cite{Balitsky:1987bk, Belyaev:1985wza}:
\begin{align}
\label{edmn13}
S_{q}(x)&= S_q^{free}(x) 
-\frac {i g_s }{16 \pi^2 x^2} \int_0^1 du \, G^{\mu \nu} (ux)
\bigg[\bar u \rlap/{x} 
\sigma_{\mu \nu} + u \sigma_{\mu \nu} \rlap/{x}
 \bigg],\\
%
S_{Q}(x)&=S_Q^{free}(x)
-i\frac{m_{Q}\,g_{s} }{16\pi ^{2}}  \int_0^1 dv \,G^{\mu \nu}(vx)\bigg[ (\sigma _{\mu \nu }{\xslash}
+{\xslash}\sigma _{\mu \nu }) 
    \frac{K_{1}\big( m_{Q}\sqrt{-x^{2}}\big) }{\sqrt{-x^{2}}}
 +2\sigma_{\mu \nu }K_{0}\big( m_{Q}\sqrt{-x^{2}}\big)\bigg],
 \label{edmn14}
\end{align}%
with  
\begin{align}
 S_q^{free}(x)&=\frac{1}{2 \pi x^2}\Big(i \frac{\xslash}{x^2}- \frac{m_q}{2}\Big),\\
 S_Q^{free}(x)&=\frac{m_{Q}^{2}}{4 \pi^{2}} \bigg[ \frac{K_{1}\big(m_{Q}\sqrt{-x^{2}}\big) }{\sqrt{-x^{2}}}
+i\frac{{\xslash}~K_{2}\big( m_{Q}\sqrt{-x^{2}}\big)}
{(\sqrt{-x^{2}})^{2}}\bigg],
\end{align}
where $G^{\mu\nu}(x)$ is the gluon field-strength tensor, and $K_i$'s are the Bessel functions.  

The correlation functions in Eqs. (\ref{eq:QCDSide})-(\ref{eq:QCDSide2}) receive both perturbative, i.e., once the photon interacts perturbatively with light and heavy quark propagators, and non-perturbative, i.e., the photon interacts with light quarks at a large distance, contributions.

To determine the nature of the perturbative contributions it is essential to undertake the subsequent replacement under the methodology characterized as: 
\begin{align}
\label{free}
S_{Q(q)}^{free}(x) \longrightarrow \int d^4z\, S_{Q(q)}^{free} (x-z)\,\rlap/{\!A}(z)\, S_{Q(q)}^{free} (z)\,.
\end{align}
where the rest of the propagators are taken into account as free propagators. This amounts to taking $\bar T_4^{\gamma} (\underline{\mu}) = 0$ and $S_{\gamma} (\underline {\mu}) = \delta(\mu_{\bar q})\delta(\mu_{q})$ as the light-cone distribution amplitude in the three particle distribution amplitudes (see Ref. \cite{Li:2020rcg}).  

To encompass non-perturbative contributions in the analysis, it is useful to apply the following formula:
 \begin{align}
\label{edmn21}
S_{q,\mu\nu}^{ab}(x) \longrightarrow -\frac{1}{4} \big[\bar{q}^a(x) \Gamma_i q^b(0)\big]\big(\Gamma_i\big)_{\mu\nu},
\end{align}
where $\Gamma_i = \{\textbf{1}$, $\gamma_5$, $\gamma_\mu$, $i\gamma_5 \gamma_\mu$, $\sigma_{\mu\nu}/2\}$.  
After the aforementioned light-quark replacement, the rest of the propagators are considered to be full propagators.

The incorporation of non-perturbative contributions into the analysis gives rise to the emergence of matrix elements such as $\langle \gamma(q)\vel \bar{q}(x) \Gamma_i G_{\mu\nu}q(0) \ver 0\rangle$ and $\langle \gamma(q)\vel \bar{q}(x) \Gamma_i q(0) \ver 0\rangle$. These matrix elements are characterized by the photon distribution amplitudes (DAs)~\cite{Ball:2002ps}. 
In the context of the current study, it is of significance to acknowledge that the photon DAs employed encompass solely contributions from light quarks. Nevertheless, it is theoretically possible for a photon to be emitted from heavy quarks over a long distance. However, the probability of long-distance photon emission from heavy quarks is significantly suppressed because of the large mass of these quarks. Such contributions are not considered within the framework of this analysis. As elucidated in Eq.~(\ref{free}), solely the short-distance photon emission from heavy quarks is taken into account. It is therefore not feasible to consider DAs encompassing heavy quarks within the context of our analysis. 
A comprehensive description of the procedures employed to encompass both perturbative and non-perturbative contributions within the calculations can be found in Refs.~\cite{Ozdem:2022vip, Ozdem:2022eds}. Once the aforementioned modifications have been performed, namely when both perturbative and non-perturbative contributions are taken into account in the analysis, the QCD representation of the correlation function is obtained.

\subsection{LCSR for magnetic moments}

Utilizing dispersion relations that consider the coefficients of the same Lorentz structures the results obtained by performing calculations on both sides of the correlation function are compared. 
In the last step, we carry out Borel transformation on the variables $-p^2$ and $-(p+q)^2$ to dominate contributions from the continuum and the higher states and boost ground states to get 
\begin{equation}
    \begin{aligned}
                \mu_{Z_{\bar b c}} \,\lambda^{2}_{Z_{\bar b c}}  e^{-\frac{m_1^2}{M_1^2}}e^{-\frac{m_2^2}{M_2^2}} &= \int_0^\infty ds_1 \int_0^\infty ds_2 \, e^{-\frac{s_1}{M_1^2}-\frac{s_2}{M_2^2}}\rho(s_1,s_2).\\
    \end{aligned}
    \label{correlated}
\end{equation}
where $m_1(m_2)$, $s_1(s_2)$ and $M_1^2(M_2^2)$ are the mass, continuum threshold and Borel parameter for the initial(final) $Z_{\bar b c}$ tetraquarks, respectively.

 To acquire the magnetic moment within the QCD light-cone sum rules, the contributions from the higher states and the continuum have been extracted utilizing quark-hadron duality ansatz: 
\begin{equation}
    \rho(s_1,s_2) \simeq \rho^{OPE}(s_1,s_2) ~~\mbox{if}~~ (s_1,s_2) \not\in {\mathbb D},
\end{equation}
where $\mathbb D$ is a domain in the $(s_1,s_2)$ plane. 
Generally, the domain $\mathbb D$ is a rectangular region defined by $s_1<s_{10}$ and $s_2<s_{20}$ for some constants $s_{10}$ and $s_{20}$, or a triangular region. 
In the present study, for brevity, continuum subtraction is performed 
via selecting $\mathbb D$ as the region determined as $s \equiv s_1 u_0 + s_2 \bar u_0 < s_0$ where $u_0\equiv \frac{M_2^2}{M_1^2+M_2^2}$ and $\bar u_0 = 1 - u_0$ . Defining  a second variable $u=\frac{s_1 u_0}{s}$, the integral in the $(s_1,s_2)$ plane can be defined as:
\begin{equation}
    \int_0^\infty ds_1 \int_0^\infty ds_2 \,e^{-\frac{s_1}{M_1^2}-\frac{s_2}{M_2^2}}\, \rho(s_1,s_2) = \int_0^\infty ds \,\rho(s)\, e^{-\frac{s}{M^2}},
\end{equation}
where,   
\begin{equation}
{\rm{M^2}}= \frac{M_1^2 M_2^2}{M_1^2+M_2^2}, ~\rm{and}  ~
    \rho(s) = \frac{s}{u_0 \bar u_0} \int_0^1 du \rho\left( s\frac{u}{u_0}, s \frac{\bar u}{\bar u_0} \right). 
\end{equation}

In the problem under review, the masses of the initial and final state tetraquarks are identical, hence we can set $M_1^2 = M_2^2 = 2M^2$, which yields $u_0=\frac{1}{2}$. Following the conclusion of the aforementioned procedure, the continuum subtraction by setting the upper limit to $s_0$  is equivalent (in the original double spectral density) to subtract everything outside the triangular region $s = s_1 u_0 + s_2 \bar u_0 \equiv s_0$.

The aforementioned processes lead to the conclusion that the magnetic moments of the $Z_{\bar b c}$ tetraquark states can be expressed in the following sum rules:
\begin{align}
\label{jmu1}
 \mu_{Z_{\bar b c}}^{J_\mu^1}\,\lambda^{2, {J_\mu^1}}_{Z_{\bar b c}}  &= e^{\frac{m_{Z_{\bar b c}}^{2,{J_\mu^1}}}{\rm{M^2}}} \,\, \rho_1(\rm{M^2},\rm{s_0}),~~~~
 \mu_{Z_{\bar b c}}^{J_\mu^2}\, \lambda^{2, {J_\mu^2}}_{Z_{\bar b c}}  = e^{\frac{m_{Z_{\bar b c}}^{2, {J_\mu^2}}}{\rm{M^2}}} \,\, \rho_2(\rm{M^2},\rm{s_0}),\\
  \mu_{Z_{\bar b c}}^{J_\mu^3}\,  \lambda^{2,{J_\mu^3}}_{Z_{\bar b c}} &= e^{\frac{m_{Z_{\bar b c}}^{2, {J_\mu^3}}}{\rm{M^2}}} \,\, \rho_3(\rm{M^2},\rm{s_0}),~~~~
 \mu_{Z_{\bar b c}}^{J_\mu^4}\, \lambda^{2,{J_\mu^4}}_{Z_{\bar b c}}  = e^{\frac{m_{Z_{\bar b c}}^{2,{J_\mu^4}}}{\rm{M^2}}} \,\, \rho_4(\rm{M^2},\rm{s_0}).
 \end{align}
It should be noted that the $(q_\mu \varepsilon_\nu-\varepsilon_\mu q_\nu)$ structure has been selected for the magnetic moment calculations of the $Z_{\bar b c}$ tetraquark states. This choice is motivated by its superior convergence properties in the OPE. The explicit momentum factors in this structure suppress higher-dimensional condensate contributions relative to the leading perturbative term.  To quantify this advantage, we have analyzed the OPE convergence within the working Borel region (specified in the next section). The perturbative contribution dominates, while dimension-3 (quark condensate), dimension-6, and higher-dimensional terms collectively contribute less than $5\%$ of the total. This favorable hierarchy justifies the OPE truncation and ensures result stability. For comparison, preliminary tests with alternative Lorentz structures (such as $(q_\mu p_\nu-p_\mu q_\nu)$-type) showed slower convergence, with condensate contributions reaching $20-25\%$ of the total OPE. Such larger non-perturbative contributions would introduce significant systematic uncertainties from poorly determined higher-dimensional vacuum condensates. Therefore, the selected $(q_\mu \varepsilon_\nu-\varepsilon_\mu q_\nu)$ structure provides the most reliable framework for extracting the magnetic moments.
 Since the functional forms are analogous, we present a representative result below---specifically for the spectral density function $\rho_1(\rm{M^2},\rm{s_0})$. The full set of corresponding expressions is provided in Appendix \ref{appa}.

Analytical formulations for the magnetic moments of $Z_{\bar b c}$ tetraquark states are presented herein. Numerical computations of these characteristics will be provided in a subsequent section.

\end{widetext}

\section{Numerical evaluations}\label{numerical}

The numerical evaluations of QCD light-cone sum rules required for the determination of magnetic moments entail the input of several quantities, the values of which are provided in Table~\ref{inputparameter}.    A further crucial input parameter in the numerical evaluations is the form employed for the photon DAs and the wave functions. The expressions in question, along with the input parameters utilized in their explicit forms, are presented in the Appendix \ref{appb}.  
%
  \begin{table}[htp]
	\addtolength{\tabcolsep}{10pt}
	\caption{ Input parameters used in calculations.
	}
	\label{inputparameter}
\begin{tabular}{l|c|ccccc}
               \hline\hline
Parameter & Value&Unit \\
                                        \hline\hline
                                        %
$m_c$&$ 1.273 \pm 0.0046$~GeV \cite{ParticleDataGroup:2024cfk}&GeV                 \\
$m_b$&$ 4.183 \pm 0.007$ \cite{ParticleDataGroup:2024cfk}&GeV                    \\                    
$m_{{Z_{\bar b c}^{{J_\mu^{1(2)}}}}}$&$  7.30 \pm 0.08 $ \cite{Wang:2019xzt}&GeV                       \\
$m_{{Z_{\bar b c}^{J_\mu^{3(4)}}}}$&$  7.31 \pm 0.08$  \cite{Wang:2019xzt}&GeV                       \\
$f_{3\gamma} $&$ -0.0039 $    \cite{Ball:2002ps}&\,\,GeV$^2$     \\
$\langle \bar qq\rangle $&$ (-0.24 \pm 0.01)^3 $ \cite{Ioffe:2005ym} &\,\,GeV$^3$                    \\
$ \langle g_s^2G^2\rangle  $&$ 0.48 \pm 0.14 $ \cite{Narison:2018nbv} &\,\,GeV$^4$                        \\
$\lambda_{{Z_{\bar b c}^{J_\mu^{1(2)}}}}$&$  (4.82 \pm 0.71) \times 10^{-2}    $ \cite{Wang:2019xzt}&\,\,GeV$^5$  \\
$\lambda_{{Z_{\bar b c}^{J_\mu^{3(4)}}}}$&$  (5.05 \pm 0.73) \times 10^{-2}    $ \cite{Wang:2019xzt}&\,\,GeV$^5$  \\
                                      \hline\hline
 \end{tabular}
\end{table}

In addition to the above-mentioned input variables, two supplementary parameters, namely the continuum threshold parameter, denoted by $\rm{s_0}$, and the Borel mass, represented by $\rm{M^2}$, are necessary for further analysis. 
 To derive robust results from the QCD light-cone sum rules, it is desirable to identify the region where the dependence of the magnetic moments on these variables is relatively weak, the so-called "working windows".  The working windows of these supplementary parameters are defined by the standard prescriptions of the methodology employed, namely pole dominance (PC) and convergence of OPE (CVG). As indicated by QCD light-cone sum rules analysis, the CVG must be appropriately constrained to ensure OPE convergence, whereas the PC must be large enough to enhance the efficiency of the single-pole approach. These limitations are expressed through the following equations: 
\begin{align}
 \mbox{PC} &=\frac{\rho_i (\rm{M^2},\rm{s_0})}{\rho_i (\rm{M^2},\infty)},~~~~
 \mbox{CVG} (\rm{M^2}, \rm{s_0}) =\frac{\rho_i^{\rm{Dim 7}} (\rm{M^2},\rm{s_0})}{\rho_i (\rm{M^2},\rm{s_0})},
 \end{align}
 where $\rho_i^{\rm{Dim 7}} (\rm{M^2},\rm{s_0})$ is the highest dimensional term in the OPE  of $\rho_i (\rm{M^2},\rm{s_0})$. 
  \begin{table}[htb!]
	\addtolength{\tabcolsep}{6pt}
	\caption{Magnetic moments of the $Z_{\bar b c}$ tetraquark states and the related working windows of the auxiliary parameters used in the sum rules.}
	\label{table}
	\begin{center}
\begin{tabular}{lccccccc}
	   \hline\hline
	  \\
	   Currents&Tetraquarks &  $\mu$\,\,($\rm{\mu_N}$)	& $\rm{M^2}\,\,(\rm{GeV}^2)$& $\rm{s_0}\,\,(\rm{GeV}^2)$& PC\,\,($\%$) & CVG\,\,($\%$)	   \\
	   \\
	   \hline\hline
$J_\mu^1 $&$[uc]_S [\bar d \bar b]_A - [uc]_A [\bar d \bar b]_S$&   $ -2.35 \pm 0.29$      & [4.5, 5.5]& [59.0, 62.0]& [63.16, 42.57]& $\ll 1$\\
$J_\mu^2 $&$[uc]_S [\bar d \bar b]_A+[uc]_A [\bar d \bar b]_S$&  $ -2.12 \pm 0.26$          & [4.5, 5.5]& [59.0, 62.0]& [63.12, 42.62]& $ \ll 1$\\
$J_\mu^3 $&$[uc]_S [\bar u \bar b]_A -[dc]_A [\bar d \bar b]_S$&   $ -2.05 \pm 0.25$      & [4.5, 5.5]& [59.0, 62.0]& [61.82, 41.40]& $ \ll 1$\\
$J_\mu^4 $&$[uc]_S [\bar u \bar b]_A+[dc]_A [\bar d \bar b]_S$&   $ -1.85 \pm 0.23$      & [4.5, 5.5]& [59.0, 62.0]& [61.89, 41.35]& $ \ll 1$\\
	   \hline\hline
\end{tabular}
\end{center}
\end{table}

\begin{figure}[htb!]
\centering
  \includegraphics[width=0.48\textwidth]{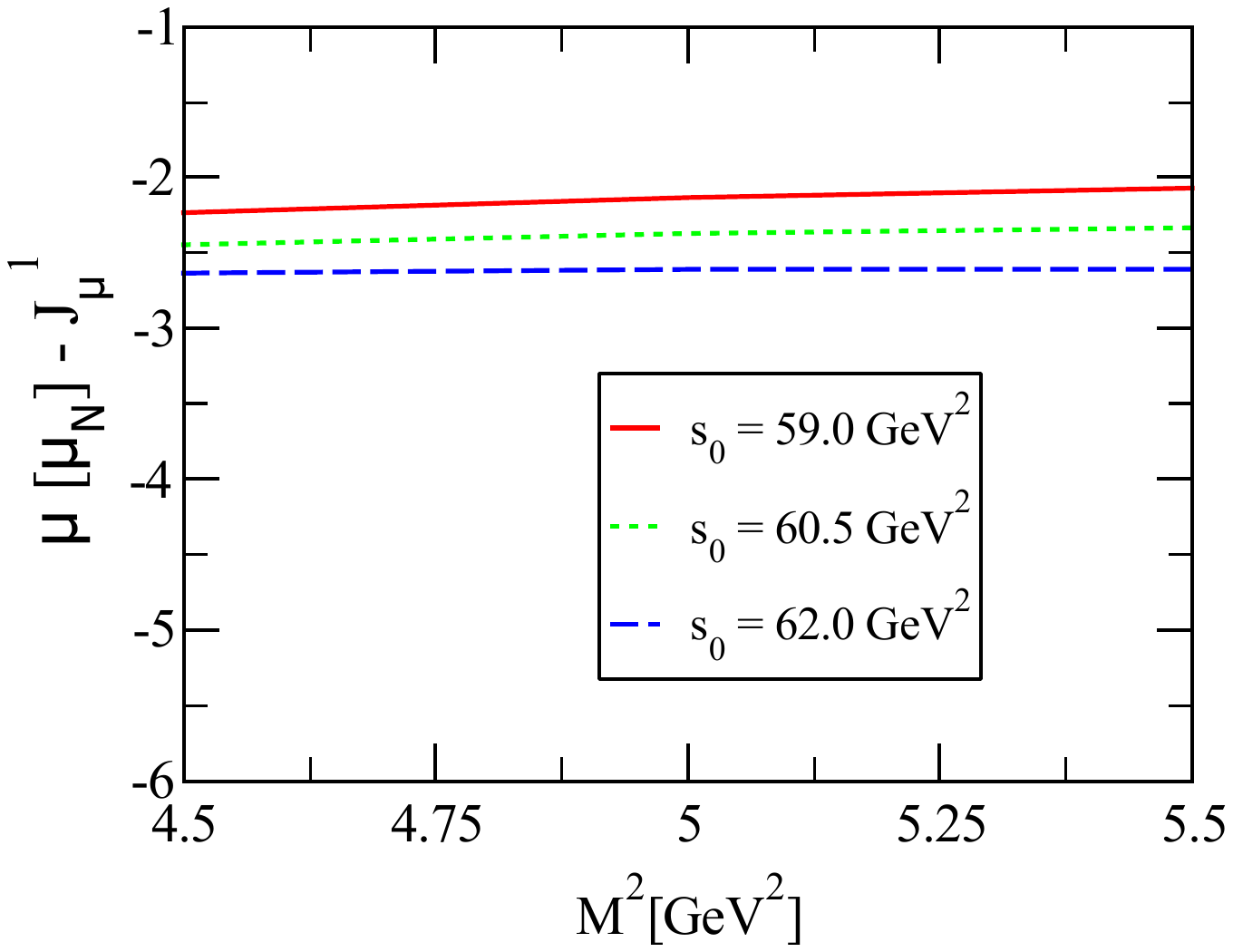} 
  \includegraphics[width=0.48\textwidth]{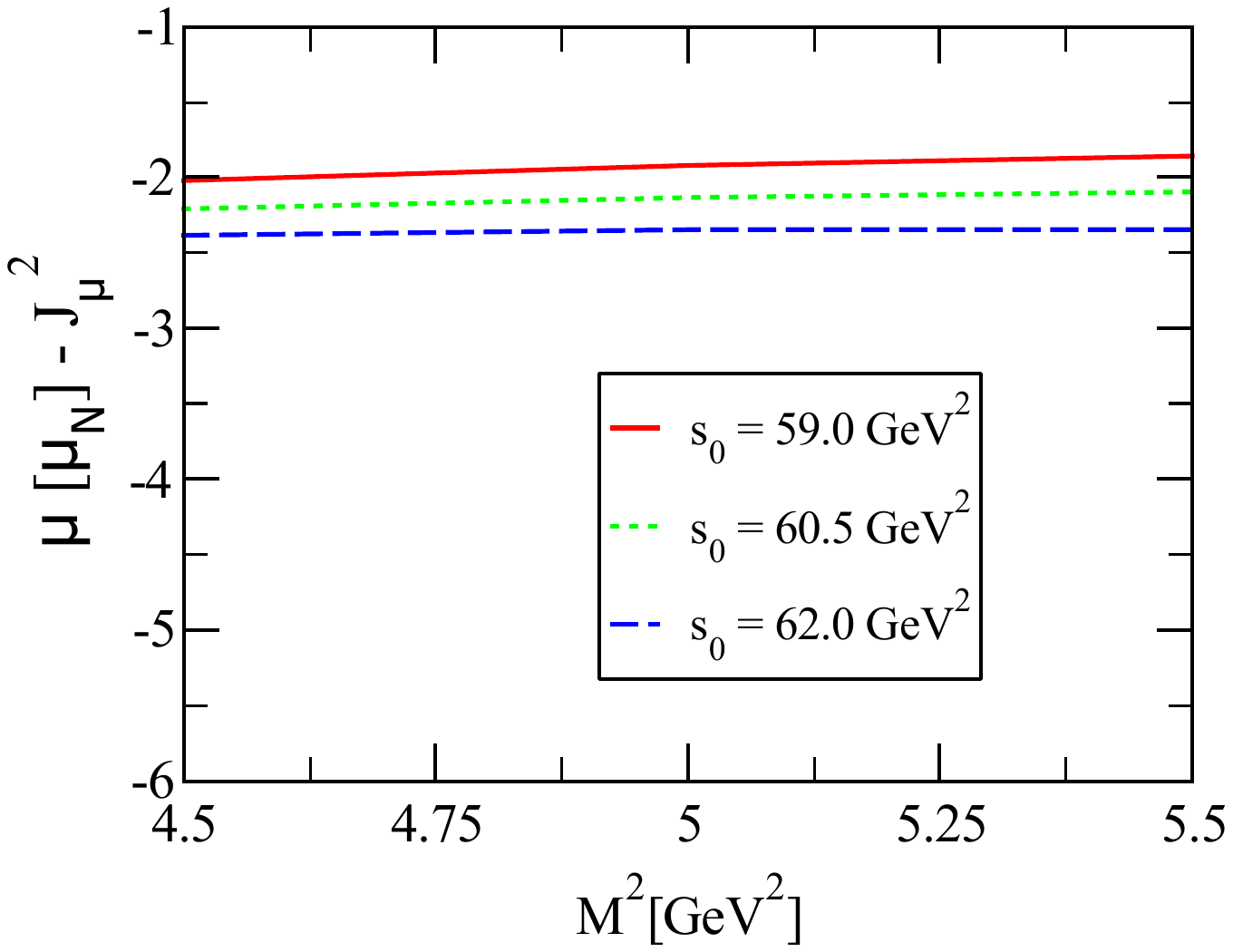} \\
  \includegraphics[width=0.48\textwidth]{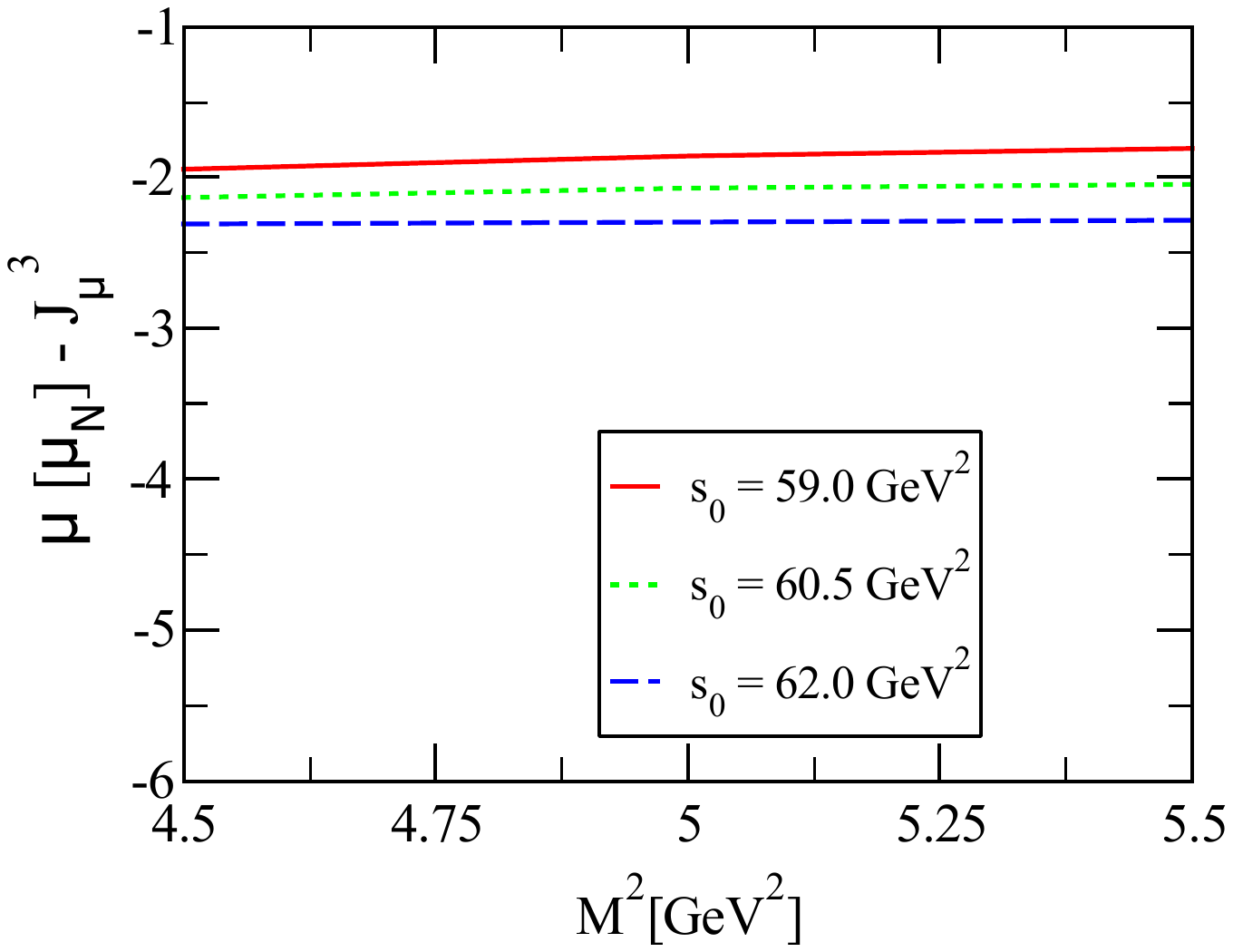} 
  \includegraphics[width=0.48\textwidth]{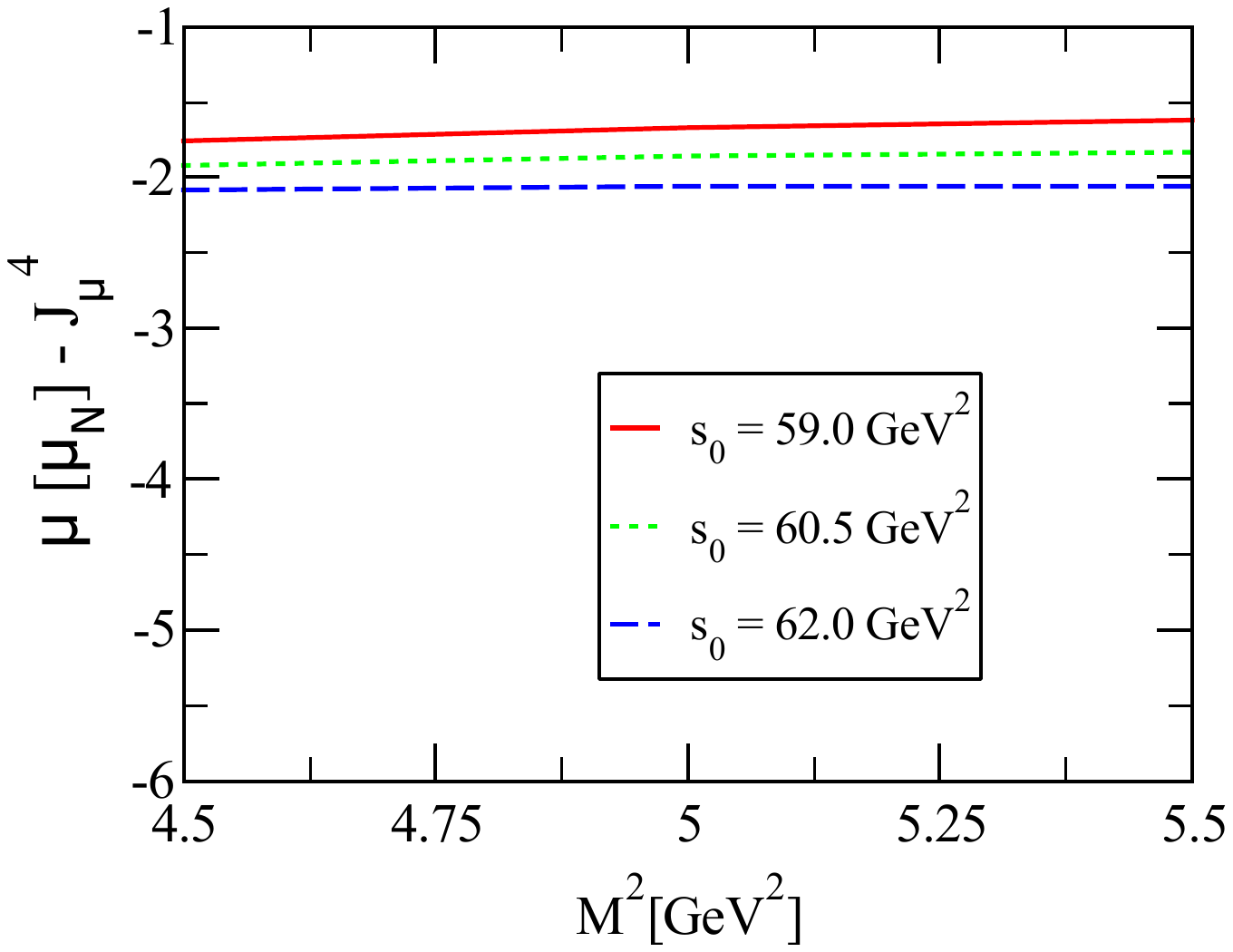}
  \caption{ Variation of magnetic moments $Z_{\bar b c}$ tetraquarks as a function of the $\rm{M^2}$ at different values of $\rm{s_0}$.}
 \label{figMsq}
  \end{figure}

 
 Under the above-mentioned requirements, the working windows of the supplementary parameters, as outlined in Table \ref{table}, are obtained. As can be seen from these results, the method satisfies the constraints.   To enhance our predictions and for completeness,  Fig.~\ref{figMsq} shows the variations in the derived magnetic moments of these states with regards to $\rm{M^2}$ and $\rm{s_0}$. As shown in the figure, the magnetic moments of these states show a relatively mild dependence on these supplementary variables. At this juncture, all criteria inherent to the QCD light-cone sum rules have been fulfilled, and we expect to make reliable predictions.

\begin{figure}[htb!]
\centering
   \subfloat[]{\includegraphics[width=0.4\textwidth]{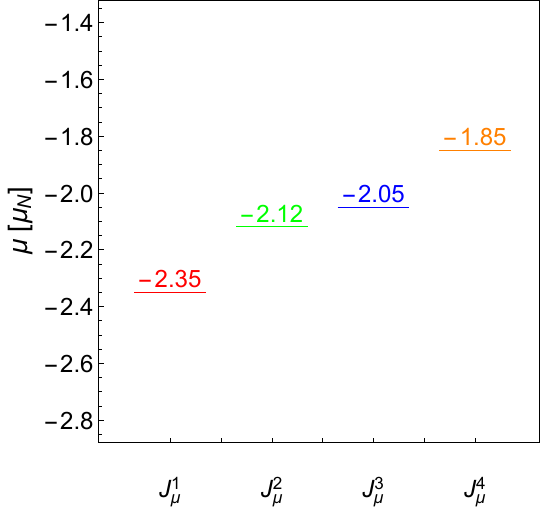}} \hspace{1.5 cm}
  \subfloat[]{\includegraphics[width=0.4\textwidth]{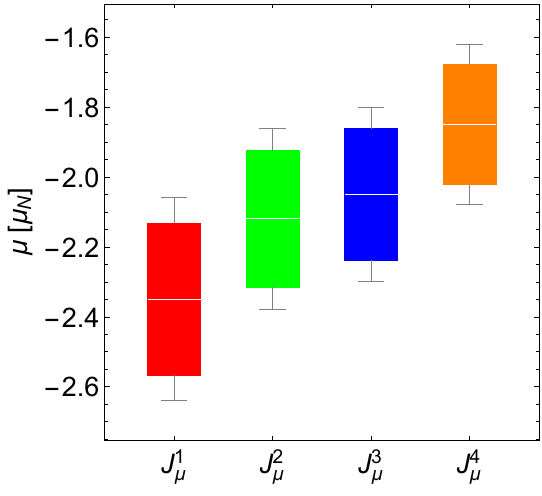}} \\
   \subfloat[]{\includegraphics[width=0.4\textwidth]{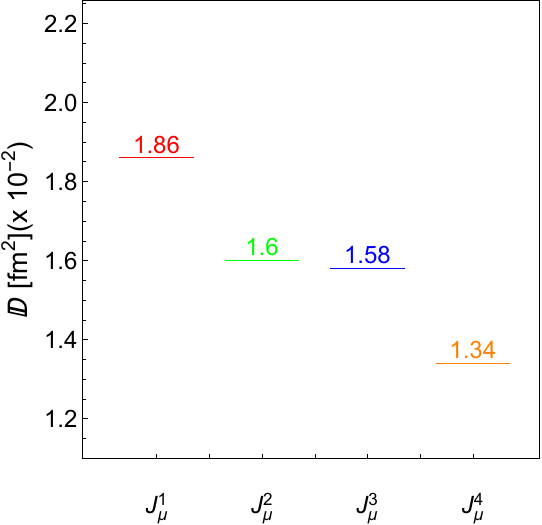}} \hspace{1.5 cm}
  \subfloat[]{\includegraphics[width=0.4\textwidth]{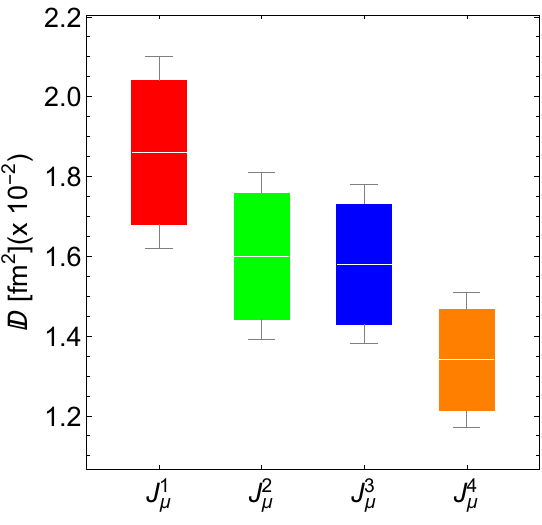}} 
  \caption{ The magnetic $(\mu)$ and quadrupole ($\mathcal D$) moments of $Z_{\bar b c}$ tetraquark states: (a) and (c) for central values, and (b) and (d) for combined with errors, respectively.}
 \label{figMsq1}
  \end{figure}

The estimated magnetic moments of the $Z_{\bar b c}$ tetraquark states, which take into account the ambiguities inherent in the input quantities and the fluctuations in the $\rm{M^2}$ and $\rm{s_0}$ working regions, have been provided in Table~\ref{table}.  Moreover, to enhance comprehension, we have provided the results of the magnetic moments, presented in conjunction with both their central values and the combined with errors, in Fig. \ref{figMsq1}.  
The uncertainties in our numerical results originate from the variations in the input parameters. The dominant sources are the continuum threshold $\mathrm{s_0}$ ($30\%$), the tetraquark residue ($20\%$), the tetraquark mass ($19\%$), and the quark masses ($12\%$). Smaller uncertainties arise from the parameters of the photon DAs ($7\%$), the Borel parameter $\mathrm{M^2}$ ($7\%$) and other inputs, such as various QCD condensates ($5\%$).

In consideration of the findings yielded by this study, the following observations have been made:
  \begin{itemize}
    
\item The main contribution of the analysis comes from the short-distance interactions of photon with quarks, which are responsible for roughly $85\%$ of the magnetic moment results.  The remaining contributions are derived from the long-distance interactions of light quarks with the photon, which is a non-perturbative contribution.  The observed decomposition of contributions---approximately 85\% from the short-distance photon-quark interaction and 15\% from long-distance effects encoded in photon DAs---is a direct consequence of the QCD sum rule framework and provides insight into the nature of the state. In the LCSR approach, the OPE is organized by operator dimension. The dominant contribution naturally arises from the leading perturbative diagram where the photon couples directly to a propagating quark, which is calculable within perturbative QCD. The substantial magnitude of this term confirms that our sum rule is sensitive primarily to the hard, partonic core of a compact tetraquark. The non-perturbative 15\%, originating from higher-twist photon DAs, incorporates soft gluon and vacuum polarization effects that dress the photon. This modest but non-negligible share is essential: it ensures the convergence of the OPE while capturing necessary hadronic cloud effects, yet its subdominant role reinforces the picture of a compact, diquark-antidiquark object rather than an extended molecular configuration. Such a balance between perturbative and non-perturbative contributions is consistent with LCSR analyses of conventional hadrons and supports the internal consistency of our calculation.

  \item The numerical results presented in Table~\ref{table1} reveal several important features regarding the internal quark dynamics of the $Z_{\bar{b}c}$ states. Firstly, the magnetic dipole moment is predominantly shaped by the contributions of the light quarks. In particular, the $u$ quark yields a large positive contribution, while the $d$ quark gives a comparably large negative one, partially canceling each other. The heavy quark contributions from the $c$ and $b$ quarks are relatively smaller in magnitude, as expected due to their larger masses, and consistently contribute with negative signs across all interpolating currents. 
  Notably, while the $u$ quark yields the largest individual positive contribution due to its charge and relatively light mass, this effect is substantially counterbalanced by the combined negative inputs of the $d$, $c$, and $b$ quarks. This cancellation indicates a delicate internal structure in which spin and spatial correlations between constituent quarks significantly influence the electromagnetic properties of the state.    The comparable contributions from $c$ and $d$ quarks, despite their $\sim 300$-fold mass difference, arise from the correlated spin structure of the tetraquark. The naive expectation $\mu \propto Q/m$ is modified by the spin expectation values $\langle S^z \rangle$ within the bound state. The light quark's large $1/m_d$ factor is compensated by its small spin matrix element in the spin-singlet diquark, while the charm quark's contribution is enhanced by its larger charge ($Q_c = +2/3$ vs $Q_d = -1/3$) and more favorable spin alignment in the wave function.
  Experimentally, such negative and moderate magnetic dipole moments could manifest in radiative transitions or influence the decay widths of the $Z_{\bar{b}c}$ states into final states involving photons. If these states are observed in future experiments, their electromagnetic multipole moments—particularly magnetic dipole transitions to nearby tetraquark or mesonic states—may serve as indirect probes of their internal spin structure. Therefore, our results may provide guidance for designing experimental searches sensitive to electromagnetic observables in the exotic hadron sector. Overall, the magnetic dipole moment emerges not merely as a static property but as a powerful probe of the interplay between quark flavor, color dynamics, and spin correlations inside exotic hadrons. The results obtained here offer both a testable prediction and a theoretical benchmark for the structure of open-flavor tetraquark states.

  \begin{table}[htb!]
	\addtolength{\tabcolsep}{6pt}
	\caption{The contribution of light and heavy quarks to the magnetic dipole moment of the $Z_{\bar b c}$ states ($\mu_N$).}
	\label{table1}
	\begin{center}
\begin{tabular}{lccccccc}
	   \hline\hline
       \\
  Currents&  $\mu_c$	& $\mu_b$& $\mu_u$&$\mu_d$ & $\mu_{tot}$ 	   
	  \\
	  \\
	   \hline\hline
$J_\mu^1 $&   $ -7.90$     &$-3.20$& $18.25$ & $-9.50$& $-2.35$\\
$J_\mu^2 $&  $ -7.10$      &$-2.88$& $16.41$ & $-8.55$& $-2.11$\\
$J_\mu^3 $&  $ -7.85$      &$-3.25$& $18.35$ & $-9.30$& $-2.05$\\
$J_\mu^4 $&  $ -7.11$      &$-2.90$& $16.45$ & $-8.29$& $-1.85$\\
	   \hline\hline
\end{tabular}
\end{center}
\end{table}

  
\item Although the masses of the states with quantum numbers $ \rm{J^{P}} = 1^{+} $  are nearly identical, there is a discrepancy in the numerical results of their magnetic moments, with a difference of approximately  $(10-15)\%$. This may assist in differentiating between quantum numbers associated with states with identical quark content.

\item The magnitude of the magnetic moment may be interpreted as indicating the experimental accessibility of the corresponding physical parameters. The magnitude of the results suggests that they may be measured in future experiments.
  The results obtained regarding the magnetic moments of the $Z_{\bar b c}$  tetraquark states may be checked within the context of different phenomenological approaches.

\item To the best of our knowledge, this is the pioneering study in the literature to investigate the magnetic moments of the $\rm{I(J^P)} = 1(1^+)$ $Z_{\bar b c}$ tetraquarks. Consequently, there is currently no available theoretical prediction or experimental data to which our numerical values can be compared. However, for illustrative purposes, we can make a comparison with the magnetic moments of $\rm{I(J^P)} = 0(1^+)$ $Z_{\bar b c}$ tetraquark states. In \cite{Ozdem:2022eds}, the magnetic moments of the $[uc] [\bar u \bar b]$ and $[dc] [\bar d \bar b]$ tetraquark states were calculated within the framework of the LCSR method, assuming a compact tetraquark picture with a $6_c \otimes \bar 6_c$ color configuration. The results obtained were $\mu_{[uc] [\bar u \bar b]} = 3.05^{+1.18}_{-0.95} \mu_N$ and $\mu_{[dc] [\bar d \bar b]}= 2.38^{+0.95}_{-0.75} \mu_N$. 
A direct comparison shows a clear numerical difference between our predictions and those of \cite{Ozdem:2022eds}. This difference originates from the fundamentally different isospin and color structures assumed in the two studies: we employ the $3_c \otimes \bar 3_c$ configuration, while \cite{Ozdem:2022eds} uses the $6_c \otimes \bar 6_c$ scheme. These distinct color configurations enforce different spin correlations within the diquark subsystems due to the Pauli exclusion principle. 
In the $3_c \otimes \bar 3_c$ model used here, the light-heavy diquark $[qc]$ is in a color antitriplet. For light quarks, this color antisymmetry typically pairs with a spin-0 (singlet) state to maintain the overall antisymmetry of the fermionic wave function. This spin-singlet structure determines specific patterns for the contributions to the magnetic moment operator. 
Conversely, in the $6_c \otimes \bar 6_c$ configuration, the diquark is in a color sextet, which is symmetric in color. To satisfy Fermi statistics, this is naturally coupled with a spin-1 (triplet) state. This change in the internal spin alignment modifies both the individual quark spin matrix elements and the interference between contributions from different quarks. Since the magnetic moment depends sensitively on the quark spin orientations as well as their charges, isospin and masses, such a structural difference logically leads to different numerical predictions. 
This comparison illustrates the sensitivity of the magnetic moment to the internal color-spin structure of tetraquarks. The significant variation between predictions based on different color configurations demonstrates that electromagnetic properties like the magnetic moment can serve as valuable discriminators between competing theoretical models of exotic hadron structure, providing complementary information to mass spectroscopy.

\item As a further consequence, the quadrupole moments of these states have also been calculated, and the results are as follows:
\begin{align}
 \mathcal{D}_{J_\mu^1}& = (1.86 \pm 0.24) \times 10^{-2}  ~\rm{fm}^2, ~~~~~~
  \mathcal{D}_{J_\mu^2} = (1.60 \pm 0.21) \times 10^{-2} ~\rm{fm}^2, \\
  \mathcal{D}_{J_\mu^3}& = (1.58 \pm 0.20) \times 10^{-2} ~\rm{fm}^2, ~~~~~~
  \mathcal{D}_{J_\mu^4} = (1.34 \pm 0.17) \times 10^{-2} ~\rm{fm}^2.
\end{align}
The quadrupole moments are small in magnitude ($|\mathcal D| \sim 0.01$--$0.02$ fm$^2$), while the magnetic moments are substantial ($|\mu| \sim 1.85$--$2.35$ $\mu_N$). The non-zero quadrupole moment results for the studied states imply a deviation from spherical charge distributions. It is well established that the geometric shape of hadrons can be inferred from the sign of their quadrupole moments: a negative value corresponds to an oblate shape, whereas a positive value indicates a prolate configuration. According to predicted results, the geometric shape of these states is prolate.
\end{itemize}

 
\section{Experimental Verification of Electromagnetic Properties}
\label{sec:experimental}

The magnetic ($\mu \sim -1.8$ to $-2.4\,\mu_N$) and quadrupole ($\mathcal D \sim 10^{-2}\,\mathrm{fm}^2$) moments calculated in this work provide quantitative predictions that can, in principle, be tested experimentally. This section discusses realistic pathways for such tests, with particular emphasis on how electromagnetic moments—central to this study—could be constrained through measurable observables.

\subsection{Production in high-energy collisions}
\label{subsec:production}

The $Z_{\bar{b}c}$ tetraquarks can be produced at both hadron and electron-positron colliders:

\begin{itemize}
    \item \textbf{LHC (LHCb):} The most promising venue. The dominant mechanism is gluon fusion $gg \to b\bar{b}c\bar{c}$ followed by hadronization. While a precise cross-section prediction requires dedicated calculations, the production of similar open-heavy-flavor states (e.g., $B_c$ mesons) is well-established at LHCb with significant yields. Given LHCb's demonstrated capability to reconstruct states with $b$- and $c$-quarks, and with projected integrated luminosities of $50\,\mathrm{fb}^{-1}$ in Run 3--4, the $Z_{\bar{b}c}$ tetraquark should be producible in sufficient numbers for discovery studies.
    
    \item \textbf{Belle II:} Cleaner events via $\Upsilon(5S)$ decays, e.g., $e^+e^- \to \Upsilon(5S) \to Z_{\bar{b}c} + \bar{D}^{(*)} + \pi$. The expected branching fractions are $\sim 10^{-5}$--$10^{-4}$, accessible with the full Belle II dataset.
\end{itemize}

\subsection{Identification through decay channels}
\label{subsec:identification}

For initial discovery, hadronic decays offer the required signal yield:
\begin{itemize}
    \item $Z_{\bar{b}c} \to B_c^+ \pi^-$ with $B_c^+ \to J/\psi(\to\mu^+\mu^-) \pi^+$
    \item $Z_{\bar{b}c} \to J/\psi(\to\mu^+\mu^-) K^-$
\end{itemize}
With $1\%$ reconstruction efficiency, $50\,\mathrm{fb}^{-1}$ at LHCb gives ${\cal O}(250)$ reconstructed events—sufficient for a $5\sigma$ observation.

\subsection{Measuring electromagnetic properties: From moments to observables}
\label{subsec:measurement}

\subsubsection{Accessing the magnetic moment via radiative transitions}

The magnetic moment can be constrained through $M1$ radiative decays between spin-partner states, e.g., $Z_{\bar{b}c}(1^+) \to Z_{\bar{b}c}(0^+) + \gamma$. The partial width is:
\begin{equation}
\Gamma_{M1} = \frac{8\alpha}{3} \frac{k_\gamma^3}{m_p^2} |\mu_{\text{trans}}|^2,
\label{eq:M1width}
\end{equation}
where $k_\gamma$ is the photon energy and $\mu_{\text{trans}}$ the transition magnetic moment. For heavy compact systems, $\mu_{\text{trans}} \approx \mu_{\text{static}}$ to a good approximation. Using our predicted $|\mu| \sim 2\,\mu_N$ and typical $k_\gamma \sim 100$ MeV, we expect $\Gamma_{M1} \sim 0.1$--$1$ keV.

\textbf{Theoretical caveat:} The approximation $\mu_{\text{trans}} \approx \mu_{\text{static}}$ neglects possible differences in the $1^+$ and $0^+$ wave functions, introducing a systematic uncertainty of $\sim 20$--$30\%$ on the extracted $\mu$ from $\Gamma_{M1}$. Full lattice QCD or model calculations of the transition form factor would be needed for a precise extraction.

\textbf{Experimental requirements:} Measuring $\Gamma_{M1}$ requires:
\begin{enumerate}
    \item Reconstruction of both the initial and final tetraquark states
    \item Detection of the soft photon ($E_\gamma \sim 50$--$300$ MeV)
    \item Sufficient statistics: Assuming $\mathcal{B}(Z^* \to Z\gamma) \sim 1\%$ and $10\%$ photon efficiency, ${\cal O}(10^4)$ reconstructed $Z$ events are needed for a $\sim 50\%$ uncertainty on $\Gamma_{M1}$
\end{enumerate}

\textbf{Experimental considerations:} Detecting $E_\gamma \sim 100$ MeV photons requires excellent calorimeter granularity and low-energy thresholds. At LHCb, converted photons or sophisticated $\pi^0$ veto algorithms would be essential for background suppression.

\subsubsection{Probing the quadrupole moment through angular distributions}

The quadrupole moment $\mathcal D$ affects the $E2/M1$ mixing ratio in radiative transitions. The photon angular distribution in $Z^* \to Z\gamma$ depends on this ratio:
\[
\frac{d\Gamma}{d\cos\theta_\gamma} \propto 1 + a_2 \cos^2\theta_\gamma,
\]
where $a_2$ is sensitive to both $\mu$ and $\mathcal D$. A precision angular analysis of cascade decays like $Z^* \to (Z \to J/\psi\pi) + \gamma$ could, in principle, extract $\mathcal D$.

\subsubsection{Additional constraints from production dynamics}

If $Z_{\bar{b}c}$ states are produced with polarization (as expected in forward LHCb acceptance), the angular distributions of decay products depend on electromagnetic form factors. While more model-dependent, such analyses provide complementary constraints.

\textbf{Alternative approaches:} If semileptonic decays $Z_{\bar{b}c} \to D^{(*)} \ell \nu$ are observed, the lepton angular distribution could provide additional constraints on the magnetic form factor, though these channels are expected to have small branching fractions.

\subsection{Feasibility assessment and challenges}
\label{subsec:feasibility}

\textbf{Required luminosity:} Constraining $\mu$ via $\Gamma_{M1}$ is statistically demanding. The need for ${\cal O}(10^4)$ reconstructed $Z$ events makes this an objective for the HL-LHC era ($\mathcal{L} \gtrsim 300\,\mathrm{fb}^{-1}$) rather than earlier runs.

\textbf{Primary experimental challenges:}
\begin{itemize}
    \item \textbf{Soft photon reconstruction:} $M1$ photons are soft ($E_\gamma \sim 50$--$300$ MeV) and must be distinguished from abundant $\pi^0 \to \gamma\gamma$ backgrounds. This requires excellent calorimeter performance and advanced particle identification.
    \item \textbf{Theoretical systematics:} Extracting the static moment $\mu$ from the measured $\Gamma_{M1}$ involves assumptions about wavefunction overlaps. The resulting $\sim 20$--$30\%$ uncertainty may dominate over statistical errors.
    \item \textbf{Background control:} Combinatorial background from conventional $B$- and $D$-meson decays could obscure both the tetraquark signals and the radiative photons.
\end{itemize}

\textbf{Realistic timeline:} Discovery via hadronic decay channels is feasible with current LHCb datasets. A first measurement of the radiative width $\Gamma_{M1}$ may be possible by the end of LHC Run 4, while precise electromagnetic moment extraction will likely require HL-LHC statistics.

\subsection{Summary and implications}
\label{subsec:summary}

This study provides specific numerical predictions for the electromagnetic properties of $Z_{\bar{b}c}$ tetraquarks. While challenging, experimental verification is grounded in established techniques: radiative transition measurements and angular analyses. A confirmed magnetic moment around $-2\,\mu_N$ would provide strong evidence for the compact diquark-antidiquark structure proposed here, offering a clear discriminant against molecular interpretations where different charge and spin distributions would likely yield different magnetic responses. The coming decade of high-luminosity heavy-flavor experiments will be crucial for testing these predictions and advancing our understanding of exotic hadron structure.

\section{Concluding notes}\label{sum}

An investigation of the electromagnetic characteristics of unconventional states may offer new insights into their internal structures. In particular, the magnetic moment characteristics may serve as a crucial physical observable for differentiating exotic states with disparate configurations or spin-parity quantum numbers. As a promising avenue for research, encompassing both opportunities and challenges, an in-depth examination of the electromagnetic properties of exotic states is crucial for advancing our understanding of unconventional states.   
Motivated by this, in this study, the magnetic moments of $ \rm{I(J^{P})} = 1(1^{+ })$ $Z_{\bar b c}$  tetraquark states are analyzed in the framework of QCD light-cone sum rules by considering the diquark-antidiquark approximation. 
Despite the nearly identical masses of the tetraquark states analyzed in this study, significant differences are observed in their magnetic moments. This may facilitate the differentiation between quantum numbers associated with states with identical quark content.  The results show that heavy quarks overcoming light quarks can determine both the sign and the magnitude of the magnetic moments of these tetraquark states.  The numerical results obtained in this study suggest that the magnetic moments of $Z_{\bar b c}$ tetraquark states may reveal aspects of their underlying structure, which could in turn distinguish between their spin-parity quantum numbers and their internal structure. The results obtained regarding the magnetic moments of the $Z_{\bar b c}$  tetraquark states may be checked within the context of different phenomenological approaches. As a further consequence, the quadrupole moments of these states have also been calculated. The quadrupole moment results obtained for these states are non-zero, indicating the presence of a non-spherical charge distribution.

The magnetic moment of hadrons constitutes an indispensable component in the calculation of the photo- and electro-production cross sections. In the future, this parameter may be derived from experimental data. As the luminosity of future runs increases, it will become possible to extract spectroscopic properties and magnetic moments of $Z_{\bar b c}$ tetraquark states from experimental facilities. This will facilitate the distinction between different theoretical configurations and contribute to a deeper understanding of the dynamics that govern their formation. Furthermore, it will be essential to determine the branching ratios of the various decay modes and decay channels of $Z_{\bar b c}$ tetraquark states.

 \appendix
 \section{The obtained sum rules for the  $J_\mu^1$ current} \label{appa}
In the Appendix, we provide the full analytical results for the magnetic moment calculations corresponding to the interpolating current  $J_\mu^1$. The explicit expressions are presented as follows:
 \begin{align}
\rho_1(\rm{M^2},\rm{s_0})&= \frac {1} {2 ^{22} \times 3^2 \times 5^2 \times 7  \pi^5}\Bigg[ 
   4 \big(9 e_d + 21 e_b - 28 e_c + 9 e_u\big) I[0, 6] + 
 3 \big(387 e_d + 126 e_b - 154 e_c + 387 e_u\big) I[1, 5]\Bigg]\nonumber\\
       & +\frac {\langle g_s^2 G^2\rangle \langle \bar q q \rangle} {2 ^{24} \times 3^4  \pi^3}\Bigg[  11 e_d \Big ((2 m_b - m_c) I_ 1[\mathcal S] - 
   2 m_b I_ 1[\mathcal T_ 1] - 4 m_b I_ 1[\mathcal T_ 2] - 
   m_c I_ 1[\mathcal T_ 2] + m_c I_ 1[\mathcal T_ 3] - 
   2 m_b I_ 1[\mathcal T_ 4] \nonumber\\
   &+ 2 m_b I_ 1[\mathcal {\tilde S}] + 
   4 m_b I_ 4[\mathcal T_ 1] + 8 m_b I_ 4[\mathcal T_ 2] + 
   2 m_c I_ 4[\mathcal T_ 2] - 2 m_c I_ 4[\mathcal T_ 3] + 
   4 m_b I_ 4[\mathcal T_ 4]\Big)+ 11 e_u \Big (m_b I_ 2[\mathcal S] + 2 m_c I_ 2[\mathcal T_ 1] \nonumber\\
   &+ 
   m_b I_ 2[\mathcal T_ 2] + 2 m_c I_ 2[\mathcal T_ 2] - 
   m_b I_ 2[\mathcal T_ 3] - 4 m_c I_ 3[\mathcal T_ 1] - 
   2 m_b I_ 3[\mathcal T_ 2] - 4 m_c I_ 3[\mathcal T_ 2] + 
   2 m_b I_ 3[\mathcal T_ 3]\Big)+32 (e_u m_b \nonumber\\
   &- e_d m_c) I_5[h_\gamma]\Bigg]\nonumber\\
       &-\frac {\langle g_s^2 G^2\rangle f_ {3 \gamma} } {2 ^{25} \times 3^6 \times 5 \pi^3}\Bigg[ 
 121 (e_d I_1[\mathcal A] - e_u I_2[\mathcal A]) I[0, 3] + 5760 (3e_d + e_u) \, m_c\,m_b \, I[0, 2] \,I_5[\psi_ {\gamma}^{\nu}]  \Bigg]\nonumber\\
   & +\frac {\langle \bar q q \rangle } {2 ^{22} \times 3^3 \times 5   \pi^3}\Bigg[2 e_d \Big (17 m_b I_ 1[\mathcal S] + 45 m_c I_ 1[\mathcal T_ 1] + 
   45 m_c I_ 1[\mathcal T_ 2] + 45 m_c I_ 1[\mathcal {\tilde  S}] - 
   94 m_b I_ 4[\mathcal S] + 72 m_b I_ 4[\mathcal T_ 1] \nonumber\\
   &- 
   171 m_c I_ 4[\mathcal T_ 1] + 72 m_b I_ 4[\mathcal T_ 2] - 
   171 m_c I_ 4[\mathcal T_ 2] - 72 m_b I_ 4[\mathcal T_ 3] - 
   72 m_b I_ 4[\mathcal T_ 4] + 72 m_b I_ 4[\mathcal {\tilde S}]\Big) \nonumber\\
  %
        &+
   e_u \Big (-11 m_c I_ 2[\mathcal S] + 
   2 \big (9 m_b I_ 2[\mathcal T_ 1] + 9 m_b I_ 2[\mathcal T_ 2] + 
       9 m_b I_ 2[\mathcal {\tilde S}] - 22 m_c I_ 3[\mathcal S] - 
       54 m_b I_ 3[\mathcal T_ 1] - 
       54 m_b I_ 3[\mathcal T_ 2]\big)\Big)
   \nonumber\\
   &
   +288 e_d m_b I_5[h_\gamma]
   \Bigg] I[0,4]\nonumber\\
    &+\frac {f_{3\gamma}} {2 ^{26} \times 3 \times 5^2 \times 7  \pi^3}\Bigg[ 1120  m_b m_c \big (4 e_d I_ 1[\mathcal A] + 
    e_u I_ 2[\mathcal A]\big) I[0, 
   4] + \big (-448 e_d I_ 1[\mathcal A] + 505 e_d I_ 1[\mathcal V] + 
    279 e_u I_ 2[\mathcal V]\nonumber\\
    &- 9792 e_d I_ 5[\psi_\gamma^\nu]\big) I[
   0, 5]\Bigg],
        \label{app1}
\end{align}
where the $I[n,m]$, and~$I_i[\mathcal{F}]$ functions are expressed as:
\begin{align}
 I[n,m]&= \int_{\mathcal M}^{\rm{s_0}} ds ~ e^{-s/\rm{M^2}}~
 s^n\,(s-\mathcal M)^m,\nonumber\\
 I_1[\mathcal{F}]&=\int D_{\alpha_i} \int_0^1 dv~ \mathcal{F}(\alpha_{\bar q},\alpha_q,\alpha_g)
 \delta'(\alpha_ q +\bar v \alpha_g-u_0),\nonumber
 \end{align}
 \begin{align}
  I_2[\mathcal{F}]&=\int D_{\alpha_i} \int_0^1 dv~ \mathcal{F}(\alpha_{\bar q},\alpha_q,\alpha_g)
 \delta'(\alpha_{\bar q}+ v \alpha_g-u_0),\nonumber\\
   I_3[\mathcal{F}]&=\int D_{\alpha_i} \int_0^1 dv~ \mathcal{F}(\alpha_{\bar q},\alpha_q,\alpha_g)
 \delta(\alpha_ q +\bar v \alpha_g-u_0),\nonumber\\
   I_4[\mathcal{F}]&=\int D_{\alpha_i} \int_0^1 dv~ \mathcal{F}(\alpha_{\bar q},\alpha_q,\alpha_g)
 \delta(\alpha_{\bar q}+ v \alpha_g-u_0),\nonumber\\
 I_5[\mathcal{F}]&=\int_0^1 du~ \mathcal{F}(u),\nonumber
 \end{align}
 where  $\mathcal M = (m_c+m_b)^2$, and $\mathcal{F}$ denotes the relevant DAs of the photon.
  
 \section{ Distribution amplitudes of the photon }\label{appb}
In this Appendix, we provide the matrix elements \( \langle \gamma(q) | \bar{q}(x) \Gamma_i q(0) | 0 \rangle \) and \( \langle \gamma(q) | \bar{q}(x) \Gamma_i G_{\mu\nu} q(0) | 0 \rangle \), which are associated with the photon distribution amplitudes (DAs), as derived in \cite{Ball:2002ps}:
\begin{eqnarray*}
\label{esbs14}
&&\langle \gamma(q) \vert  \bar q(x) \gamma_\mu q(0) \vert 0 \rangle
= e_q f_{3 \gamma} \left(\varepsilon_\mu - q_\mu \frac{\varepsilon
x}{q x} \right) \int_0^1 du e^{i \bar u q x} \psi^v(u)
\nonumber \\
&&\langle \gamma(q) \vert \bar q(x) \gamma_\mu \gamma_5 q(0) \vert 0
\rangle  = - \frac{1}{4} e_q f_{3 \gamma} \epsilon_{\mu \nu \alpha
\beta } \varepsilon^\nu q^\alpha x^\beta \int_0^1 du e^{i \bar u q
x} \psi^a(u)
\nonumber \\
&&\langle \gamma(q) \vert  \bar q(x) \sigma_{\mu \nu} q(0) \vert  0
\rangle  = -i e_q \langle \bar q q \rangle (\varepsilon_\mu q_\nu - \varepsilon_\nu
q_\mu) \int_0^1 du e^{i \bar u qx} \left(\chi \varphi_\gamma(u) +
\frac{x^2}{16} \mathbb{A}  (u) \right) \nonumber \\ 
&&-\frac{i}{2(qx)}  e_q \bar qq \left[x_\nu \left(\varepsilon_\mu - q_\mu
\frac{\varepsilon x}{qx}\right) - x_\mu \left(\varepsilon_\nu -
q_\nu \frac{\varepsilon x}{q x}\right) \right] \int_0^1 du e^{i \bar
u q x} h_\gamma(u)
\nonumber \\
&&\langle \gamma(q) | \bar q(x) g_s G_{\mu \nu} (v x) q(0) \vert 0
\rangle = -i e_q \langle \bar q q \rangle \left(\varepsilon_\mu q_\nu - \varepsilon_\nu
q_\mu \right) \int {\cal D}\alpha_i e^{i (\alpha_{\bar q} + v
\alpha_g) q x} {\cal S}(\alpha_i)
\nonumber \\
&&\langle \gamma(q) | \bar q(x) g_s \tilde G_{\mu \nu}(v
x) i \gamma_5  q(0) \vert 0 \rangle = -i e_q \langle \bar q q \rangle \left(\varepsilon_\mu q_\nu -
\varepsilon_\nu q_\mu \right) \int {\cal D}\alpha_i e^{i
(\alpha_{\bar q} + v \alpha_g) q x} \tilde {\cal S}(\alpha_i)
\nonumber \\
&&\langle \gamma(q) \vert \bar q(x) g_s \tilde G_{\mu \nu}(v x)
\gamma_\alpha \gamma_5 q(0) \vert 0 \rangle = e_q f_{3 \gamma}
q_\alpha (\varepsilon_\mu q_\nu - \varepsilon_\nu q_\mu) \int {\cal
D}\alpha_i e^{i (\alpha_{\bar q} + v \alpha_g) q x} {\cal
A}(\alpha_i)
\nonumber \\
&&\langle \gamma(q) \vert \bar q(x) g_s G_{\mu \nu}(v x) i
\gamma_\alpha q(0) \vert 0 \rangle = e_q f_{3 \gamma} q_\alpha
(\varepsilon_\mu q_\nu - \varepsilon_\nu q_\mu) \int {\cal
D}\alpha_i e^{i (\alpha_{\bar q} + v \alpha_g) q x} {\cal
V}(\alpha_i) \nonumber\\
&& \langle \gamma(q) \vert \bar q(x)
\sigma_{\alpha \beta} g_s G_{\mu \nu}(v x) q(0) \vert 0 \rangle  =
e_q \langle \bar q q \rangle \left\{
        \left[\left(\varepsilon_\mu - q_\mu \frac{\varepsilon x}{q x}\right)\left(g_{\alpha \nu} -
        \frac{1}{qx} (q_\alpha x_\nu + q_\nu x_\alpha)\right) \right. \right. q_\beta
\nonumber \\
 && -
         \left(\varepsilon_\mu - q_\mu \frac{\varepsilon x}{q x}\right)\left(g_{\beta \nu} -
        \frac{1}{qx} (q_\beta x_\nu + q_\nu x_\beta)\right) q_\alpha
-
         \left(\varepsilon_\nu - q_\nu \frac{\varepsilon x}{q x}\right)\left(g_{\alpha \mu} -
        \frac{1}{qx} (q_\alpha x_\mu + q_\mu x_\alpha)\right) q_\beta
\nonumber \\
&&+
         \left. \left(\varepsilon_\nu - q_\nu \frac{\varepsilon x}{q.x}\right)\left( g_{\beta \mu} -
        \frac{1}{qx} (q_\beta x_\mu + q_\mu x_\beta)\right) q_\alpha \right]
   \int {\cal D}\alpha_i e^{i (\alpha_{\bar q} + v \alpha_g) qx} {\cal T}_1(\alpha_i)
\nonumber \\
 &&+
        \left[\left(\varepsilon_\alpha - q_\alpha \frac{\varepsilon x}{qx}\right)
        \left(g_{\mu \beta} - \frac{1}{qx}(q_\mu x_\beta + q_\beta x_\mu)\right) \right. q_\nu
\nonumber \\
&&-
         \left(\varepsilon_\alpha - q_\alpha \frac{\varepsilon x}{qx}\right)
        \left(g_{\nu \beta} - \frac{1}{qx}(q_\nu x_\beta + q_\beta x_\nu)\right)  q_\mu
\nonumber \\
  && -
         \left(\varepsilon_\beta - q_\beta \frac{\varepsilon x}{qx}\right)
        \left(g_{\mu \alpha} - \frac{1}{qx}(q_\mu x_\alpha + q_\alpha x_\mu)\right) q_\nu
\nonumber \\ 
&&+
         \left. \left(\varepsilon_\beta - q_\beta \frac{\varepsilon x}{qx}\right)
        \left(g_{\nu \alpha} - \frac{1}{qx}(q_\nu x_\alpha + q_\alpha x_\nu) \right) q_\mu
        \right]      
    \int {\cal D} \alpha_i e^{i (\alpha_{\bar q} + v \alpha_g) qx} {\cal T}_2(\alpha_i)
\nonumber 
\end{eqnarray*}
\begin{eqnarray*}
&&+\frac{1}{qx} (q_\mu x_\nu - q_\nu x_\mu)
        (\varepsilon_\alpha q_\beta - \varepsilon_\beta q_\alpha)
    \int {\cal D} \alpha_i e^{i (\alpha_{\bar q} + v \alpha_g) qx} {\cal T}_3(\alpha_i)
\nonumber \\ &&+
        \left. \frac{1}{qx} (q_\alpha x_\beta - q_\beta x_\alpha)
        (\varepsilon_\mu q_\nu - \varepsilon_\nu q_\mu)
    \int {\cal D} \alpha_i e^{i (\alpha_{\bar q} + v \alpha_g) qx} {\cal T}_4(\alpha_i)
                        \right\},
\end{eqnarray*}
where $\varphi_\gamma(u)$ is the DA of leading twist-2, $\psi^v(u)$,
$\psi^a(u)$, ${\cal A}(\alpha_i)$ and ${\cal V}(\alpha_i)$, are the twist-3 amplitudes, and
$h_\gamma(u)$, $\mathbb{A}(u)$, ${\cal S}(\alpha_i)$, ${\cal{\tilde S}}(\alpha_i)$, ${\cal T}_1(\alpha_i)$, ${\cal T}_2(\alpha_i)$, ${\cal T}_3(\alpha_i)$ 
and ${\cal T}_4(\alpha_i)$ are the
twist-4 photon DAs.
The measure ${\cal D} \alpha_i$ is defined as
\begin{eqnarray*}
\label{nolabel05}
\int {\cal D} \alpha_i = \int_0^1 d \alpha_{\bar q} \int_0^1 d
\alpha_q \int_0^1 d \alpha_g \delta(1-\alpha_{\bar
q}-\alpha_q-\alpha_g)~.\nonumber
\end{eqnarray*}

The forms of the DAs that are incorporated into the matrix elements above are given by:
\begin{eqnarray}
\varphi_\gamma(u) &=& 6 u \bar u \left( 1 + \varphi_2(\mu)
C_2^{\frac{3}{2}}(u - \bar u) \right),
\nonumber \\
\psi^v(u) &=& 3 \left(3 (2 u - 1)^2 -1 \right)+\frac{3}{64} \left(15
w^V_\gamma - 5 w^A_\gamma\right)
                        \left(3 - 30 (2 u - 1)^2 + 35 (2 u -1)^4
                        \right),
\nonumber \\
\psi^a(u) &=& \left(1- (2 u -1)^2\right)\left(5 (2 u -1)^2 -1\right)
\frac{5}{2}
    \left(1 + \frac{9}{16} w^V_\gamma - \frac{3}{16} w^A_\gamma
    \right),
\nonumber \\
h_\gamma(u) &=& - 10 \left(1 + 2 \kappa^+\right) C_2^{\frac{1}{2}}(u
- \bar u),
\nonumber \\
\mathbb{A}(u) &=& 40 u^2 \bar u^2 \left(3 \kappa - \kappa^+
+1\right)  +
        8 (\zeta_2^+ - 3 \zeta_2) \left[u \bar u (2 + 13 u \bar u) \right.
\nonumber \\ && + \left.
                2 u^3 (10 -15 u + 6 u^2) \ln(u) + 2 \bar u^3 (10 - 15 \bar u + 6 \bar u^2)
        \ln(\bar u) \right],
\nonumber \\
{\cal A}(\alpha_i) &=& 360 \alpha_q \alpha_{\bar q} \alpha_g^2
        \left(1 + w^A_\gamma \frac{1}{2} (7 \alpha_g - 3)\right),
\nonumber \\
{\cal V}(\alpha_i) &=& 540 w^V_\gamma (\alpha_q - \alpha_{\bar q})
\alpha_q \alpha_{\bar q}
                \alpha_g^2,
\nonumber \\
{\cal T}_1(\alpha_i) &=& -120 (3 \zeta_2 + \zeta_2^+)(\alpha_{\bar
q} - \alpha_q)
        \alpha_{\bar q} \alpha_q \alpha_g,
\nonumber \\
{\cal T}_2(\alpha_i) &=& 30 \alpha_g^2 (\alpha_{\bar q} - \alpha_q)
    \left((\kappa - \kappa^+) + (\zeta_1 - \zeta_1^+)(1 - 2\alpha_g) +
    \zeta_2 (3 - 4 \alpha_g)\right),
\nonumber \\
{\cal T}_3(\alpha_i) &=& - 120 (3 \zeta_2 - \zeta_2^+)(\alpha_{\bar
q} -\alpha_q)
        \alpha_{\bar q} \alpha_q \alpha_g,
\nonumber \\
{\cal T}_4(\alpha_i) &=& 30 \alpha_g^2 (\alpha_{\bar q} - \alpha_q)
    \left((\kappa + \kappa^+) + (\zeta_1 + \zeta_1^+)(1 - 2\alpha_g) +
    \zeta_2 (3 - 4 \alpha_g)\right),\nonumber \\
{\cal S}(\alpha_i) &=& 30\alpha_g^2\{(\kappa +
\kappa^+)(1-\alpha_g)+(\zeta_1 + \zeta_1^+)(1 - \alpha_g)(1 -
2\alpha_g)\nonumber +\zeta_2[3 (\alpha_{\bar q} - \alpha_q)^2-\alpha_g(1 - \alpha_g)]\},\nonumber \\
\tilde {\cal S}(\alpha_i) &=&-30\alpha_g^2\{(\kappa -\kappa^+)(1-\alpha_g)+(\zeta_1 - \zeta_1^+)(1 - \alpha_g)(1 -
2\alpha_g)\nonumber +\zeta_2 [3 (\alpha_{\bar q} -\alpha_q)^2-\alpha_g(1 - \alpha_g)]\},
\end{eqnarray}

\noindent where $\varphi_2(1~GeV) = 0$, 
$w^V_\gamma = 3.8 \pm 1.8$, $w^A_\gamma = -2.1 \pm 1.0$, $\kappa = 0.2$, $\kappa^+ = 0$, $\zeta_1 = 0.4$, and $\zeta_2 = 0.3$.

\bibliography{ZbbarcMDMIsopinv2.bib}

@article{Belle:2003nnu,
    author = "Choi, S. K. and others",
    collaboration = "Belle",
    title = "{Observation of a narrow charmonium-like state in exclusive $B^\pm \to K^\pm \pi^+ \pi^- J/\psi$ decays}",
    eprint = "hep-ex/0309032",
    archivePrefix = "arXiv",
    doi = "10.1103/PhysRevLett.91.262001",
    journal = "Phys. Rev. Lett.",
    volume = "91",
    pages = "262001",
    year = "2003"
}

@article{Esposito:2014rxa,
    author = "Esposito, Angelo and Guerrieri, Andrea L. and Piccinini, Fulvio and Pilloni, Alessandro and Polosa, Antonio D.",
    title = "{Four-Quark Hadrons: an Updated Review}",
    eprint = "1411.5997",
    archivePrefix = "arXiv",
    primaryClass = "hep-ph",
    doi = "10.1142/S0217751X15300021",
    journal = "Int. J. Mod. Phys. A",
    volume = "30",
    pages = "1530002",
    year = "2015"
}

@article{Esposito:2016noz,
    author = "Esposito, A. and Pilloni, A. and Polosa, A. D.",
    title = "{Multiquark Resonances}",
    eprint = "1611.07920",
    archivePrefix = "arXiv",
    primaryClass = "hep-ph",
    reportNumber = "JLAB-THY-16-2301",
    doi = "10.1016/j.physrep.2016.11.002",
    journal = "Phys. Rept.",
    volume = "668",
    pages = "1--97",
    year = "2017"
}

@article{Olsen:2017bmm,
    author = "Olsen, Stephen Lars and Skwarnicki, Tomasz and Zieminska, Daria",
    title = "{Nonstandard heavy mesons and baryons: Experimental evidence}",
    eprint = "1708.04012",
    archivePrefix = "arXiv",
    primaryClass = "hep-ph",
    doi = "10.1103/RevModPhys.90.015003",
    journal = "Rev. Mod. Phys.",
    volume = "90",
    number = "1",
    pages = "015003",
    year = "2018"
}

@article{Lebed:2016hpi,
    author = "Lebed, Richard F. and Mitchell, Ryan E. and Swanson, Eric S.",
    title = "{Heavy-Quark QCD Exotica}",
    eprint = "1610.04528",
    archivePrefix = "arXiv",
    primaryClass = "hep-ph",
    doi = "10.1016/j.ppnp.2016.11.003",
    journal = "Prog. Part. Nucl. Phys.",
    volume = "93",
    pages = "143--194",
    year = "2017"
}

@article{Nielsen:2009uh,
    author = "Nielsen, Marina and Navarra, Fernando S. and Lee, Su Houng",
    title = "{New Charmonium States in QCD Sum Rules: A Concise Review}",
    eprint = "0911.1958",
    archivePrefix = "arXiv",
    primaryClass = "hep-ph",
    doi = "10.1016/j.physrep.2010.07.005",
    journal = "Phys. Rept.",
    volume = "497",
    pages = "41--83",
    year = "2010"
}

@article{Brambilla:2019esw,
    author = "Brambilla, Nora and Eidelman, Simon and Hanhart, Christoph and Nefediev, Alexey and Shen, Cheng-Ping and Thomas, Christopher E. and Vairo, Antonio and Yuan, Chang-Zheng",
    title = "{The $XYZ$ states: experimental and theoretical status and perspectives}",
    eprint = "1907.07583",
    archivePrefix = "arXiv",
    primaryClass = "hep-ex",
    reportNumber = "TUM-EFT 125/19",
    doi = "10.1016/j.physrep.2020.05.001",
    journal = "Phys. Rept.",
    volume = "873",
    pages = "1--154",
    year = "2020"
}

@article{Agaev:2020zad,
    author = "Agaev, Shahin and Azizi, Kazem and Sundu, Hayriye",
    title = "{Four-quark exotic mesons}",
    eprint = "2004.12079",
    archivePrefix = "arXiv",
    primaryClass = "hep-ph",
    doi = "10.3906/fiz-2003-15",
    journal = "Turk. J. Phys.",
    volume = "44",
    number = "2",
    pages = "95--173",
    year = "2020"
}

@article{Chen:2016qju,
    author = "Chen, Hua-Xing and Chen, Wei and Liu, Xiang and Zhu, Shi-Lin",
    title = "{The hidden-charm pentaquark and tetraquark states}",
    eprint = "1601.02092",
    archivePrefix = "arXiv",
    primaryClass = "hep-ph",
    doi = "10.1016/j.physrep.2016.05.004",
    journal = "Phys. Rept.",
    volume = "639",
    pages = "1--121",
    year = "2016"
}

@article{Ali:2017jda,
    author = {Ali, Ahmed and Lange, Jens S{\"o}ren and Stone, Sheldon},
    title = "{Exotics: Heavy Pentaquarks and Tetraquarks}",
    eprint = "1706.00610",
    archivePrefix = "arXiv",
    primaryClass = "hep-ph",
    reportNumber = "DESY-17-071",
    doi = "10.1016/j.ppnp.2017.08.003",
    journal = "Prog. Part. Nucl. Phys.",
    volume = "97",
    pages = "123--198",
    year = "2017"
}

@article{Guo:2017jvc,
    author = "Guo, Feng-Kun and Hanhart, Christoph and Mei{\ss}ner, Ulf-G. and Wang, Qian and Zhao, Qiang and Zou, Bing-Song",
    title = "{Hadronic molecules}",
    eprint = "1705.00141",
    archivePrefix = "arXiv",
    primaryClass = "hep-ph",
    doi = "10.1103/RevModPhys.90.015004",
    journal = "Rev. Mod. Phys.",
    volume = "90",
    number = "1",
    pages = "015004",
    year = "2018",
    note = "[Erratum: Rev.Mod.Phys. 94, 029901 (2022)]"
}

@article{Liu:2019zoy,
    author = "Liu, Yan-Rui and Chen, Hua-Xing and Chen, Wei and Liu, Xiang and Zhu, Shi-Lin",
    title = "{Pentaquark and Tetraquark states}",
    eprint = "1903.11976",
    archivePrefix = "arXiv",
    primaryClass = "hep-ph",
    doi = "10.1016/j.ppnp.2019.04.003",
    journal = "Prog. Part. Nucl. Phys.",
    volume = "107",
    pages = "237--320",
    year = "2019"
}

@article{Yang:2020atz,
    author = "Yang, Gang and Ping, Jialun and Segovia, Jorge",
    title = "{Tetra- and penta-quark structures in the constituent quark model}",
    eprint = "2009.00238",
    archivePrefix = "arXiv",
    primaryClass = "hep-ph",
    doi = "10.3390/sym12111869",
    journal = "Symmetry",
    volume = "12",
    number = "11",
    pages = "1869",
    year = "2020"
}

@article{Dong:2021juy,
    author = "Dong, Xiang-Kun and Guo, Feng-Kun and Zou, Bing-Song",
    title = "{A survey of heavy-antiheavy hadronic molecules}",
    eprint = "2101.01021",
    archivePrefix = "arXiv",
    primaryClass = "hep-ph",
    doi = "10.13725/j.cnki.pip.2021.02.001",
    journal = "Progr. Phys.",
    volume = "41",
    pages = "65--93",
    year = "2021"
}

@article{Dong:2021bvy,
    author = "Dong, Xiang-Kun and Guo, Feng-Kun and Zou, Bing-Song",
    title = "{A survey of heavy{\textendash}heavy hadronic molecules}",
    eprint = "2108.02673",
    archivePrefix = "arXiv",
    primaryClass = "hep-ph",
    doi = "10.1088/1572-9494/ac27a2",
    journal = "Commun. Theor. Phys.",
    volume = "73",
    number = "12",
    pages = "125201",
    year = "2021"
}

@article{Meng:2022ozq,
    author = "Meng, Lu and Wang, Bo and Wang, Guang-Juan and Zhu, Shi-Lin",
    title = "{Chiral perturbation theory for heavy hadrons and chiral effective field theory for heavy hadronic molecules}",
    eprint = "2204.08716",
    archivePrefix = "arXiv",
    primaryClass = "hep-ph",
    doi = "10.1016/j.physrep.2023.04.003",
    journal = "Phys. Rept.",
    volume = "1019",
    pages = "1--149",
    year = "2023"
}

@article{Chen:2022asf,
    author = "Chen, Hua-Xing and Chen, Wei and Liu, Xiang and Liu, Yan-Rui and Zhu, Shi-Lin",
    title = "{An updated review of the new hadron states}",
    eprint = "2204.02649",
    archivePrefix = "arXiv",
    primaryClass = "hep-ph",
    doi = "10.1088/1361-6633/aca3b6",
    journal = "Rept. Prog. Phys.",
    volume = "86",
    number = "2",
    pages = "026201",
    year = "2023"
}

@article{Zhang:2009vs,
    author = "Zhang, Jian-Rong and Huang, Ming-Qiu",
    title = "{{Q anti-q}{anti-Q-(prime)q} molecular states}",
    eprint = "0906.0090",
    archivePrefix = "arXiv",
    primaryClass = "hep-ph",
    doi = "10.1103/PhysRevD.80.056004",
    journal = "Phys. Rev. D",
    volume = "80",
    pages = "056004",
    year = "2009"
}

@article{Zhang:2009em,
    author = "Zhang, Jian-Rong and Huang, Ming-Qiu",
    title = "{{Q anti-s}{anti-Q-(prime)s} molecular states in QCD sum rules}",
    eprint = "0905.4672",
    archivePrefix = "arXiv",
    primaryClass = "hep-ph",
    doi = "10.1088/0253-6102/54/6/22",
    journal = "Commun. Theor. Phys.",
    volume = "54",
    pages = "1075--1090",
    year = "2010"
}

@article{Sun:2012sy,
    author = "Sun, Zhi-Feng and Liu, Xiang and Nielsen, Marina and Zhu, Shi-Lin",
    title = "{Hadronic molecules with both open charm and bottom}",
    eprint = "1203.1090",
    archivePrefix = "arXiv",
    primaryClass = "hep-ph",
    doi = "10.1103/PhysRevD.85.094008",
    journal = "Phys. Rev. D",
    volume = "85",
    pages = "094008",
    year = "2012"
}

@article{Albuquerque:2012rq,
    author = "Albuquerque, Raphael M. and Liu, Xiang and Nielsen, Marina",
    title = "{Exotic $B_c$-like molecules in QCD Sum Rules}",
    eprint = "1203.6569",
    archivePrefix = "arXiv",
    primaryClass = "hep-ph",
    doi = "10.1016/j.physletb.2012.10.063",
    journal = "Phys. Lett. B",
    volume = "718",
    pages = "492--498",
    year = "2012"
}

@article{Chen:2013aba,
    author = "Chen, Wei and Steele, T. G. and Zhu, Shi-Lin",
    title = "{Exotic open-flavor $bc\bar{q}\bar{q}$, $bc\bar{s}\bar{s}$ and $qc\bar{q}\bar{b}$, $sc\bar{s}\bar{b}$ tetraquark states}",
    eprint = "1310.8337",
    archivePrefix = "arXiv",
    primaryClass = "hep-ph",
    doi = "10.1103/PhysRevD.89.054037",
    journal = "Phys. Rev. D",
    volume = "89",
    number = "5",
    pages = "054037",
    year = "2014"
}

@article{Agaev:2016dsg,
    author = "Agaev, S. S. and Azizi, K. and Sundu, H.",
    title = "{Open charm-bottom scalar tetraquarks and their strong decays}",
    eprint = "1611.00293",
    archivePrefix = "arXiv",
    primaryClass = "hep-ph",
    doi = "10.1103/PhysRevD.95.034008",
    journal = "Phys. Rev. D",
    volume = "95",
    number = "3",
    pages = "034008",
    year = "2017"
}

@article{Agaev:2017uky,
    author = "Agaev, S. S. and Azizi, K. and Sundu, H.",
    title = "{Open charm-bottom axial-vector tetraquarks and their properties}",
    eprint = "1702.08230",
    archivePrefix = "arXiv",
    primaryClass = "hep-ph",
    doi = "10.1140/epjc/s10052-017-4892-8",
    journal = "Eur. Phys. J. C",
    volume = "77",
    number = "5",
    pages = "321",
    year = "2017"
}

@article{Wang:2020jgb,
    author = "Wang, Qi-Nan and Chen, Wei",
    title = "{Fully open-flavor tetraquark states $bc\bar{q}\bar{s}$ and $sc\bar{q}\bar{b}$ with $J^{P}=0^{+},1^{+}$}",
    eprint = "2002.04243",
    archivePrefix = "arXiv",
    primaryClass = "hep-ph",
    doi = "10.1140/epjc/s10052-020-7938-2",
    journal = "Eur. Phys. J. C",
    volume = "80",
    number = "5",
    pages = "389",
    year = "2020"
}

@article{Wang:2019xzt,
    author = "Wang, Zhi-Gang",
    title = "{Analysis of the axialvector $B_c$-like tetraquark states with the QCD sum rules}",
    eprint = "1907.10921",
    archivePrefix = "arXiv",
    primaryClass = "hep-ph",
    doi = "10.1209/0295-5075/128/11001",
    journal = "EPL",
    volume = "128",
    number = "1",
    pages = "11001",
    year = "2019"
}

@article{Wu:2018xdi,
    author = "Wu, Jing and Liu, Xiang and Liu, Yan-Rui and Zhu, Shi-Lin",
    title = "{Systematic studies of charmonium-, bottomonium-, and $B_c$-like tetraquark states}",
    eprint = "1810.06886",
    archivePrefix = "arXiv",
    primaryClass = "hep-ph",
    doi = "10.1103/PhysRevD.99.014037",
    journal = "Phys. Rev. D",
    volume = "99",
    number = "1",
    pages = "014037",
    year = "2019"
}

@article{Ortega:2020uvc,
    author = "Ortega, Pablo G. and Segovia, Jorge and Entem, David R. and Fernandez, Francisco",
    title = "{Spectroscopy of $\mathbf {B_c}$ mesons and the possibility of finding exotic $\mathbf {B_c}$-like structures}",
    eprint = "2001.08093",
    archivePrefix = "arXiv",
    primaryClass = "hep-ph",
    doi = "10.1140/epjc/s10052-020-7764-6",
    journal = "Eur. Phys. J. C",
    volume = "80",
    number = "3",
    pages = "223",
    year = "2020"
}

@article{Ozdem:2024txt,
    author = {{\"O}zdem, Ula{\c{s}}},
    title = "{Investigating the underlying structure of vector hidden-charm tetraquark states via their electromagnetic characteristics}",
    eprint = "2412.06447",
    archivePrefix = "arXiv",
    primaryClass = "hep-ph",
    doi = "10.1103/PhysRevD.111.054009",
    journal = "Phys. Rev. D",
    volume = "111",
    number = "5",
    pages = "054009",
    year = "2025"
}

@article{Ozdem:2024dbq,
    author = {{\"O}zdem, Ula{\c{s}}},
    title = "{Unveiling the underlying structure of axial-vector bottom-charm tetraquarks in the light of their magnetic moments}",
    eprint = "2403.16191",
    archivePrefix = "arXiv",
    primaryClass = "hep-ph",
    doi = "10.1007/JHEP05(2024)301",
    journal = "JHEP",
    volume = "05",
    pages = "301",
    year = "2024"
}

@article{Ozdem:2024lpk,
    author = {{\"O}zdem, U. and Azizi, K.},
    title = "{Electromagnetic properties of vector doubly charmed tetraquark states}",
    eprint = "2401.04798",
    archivePrefix = "arXiv",
    primaryClass = "hep-ph",
    doi = "10.1103/PhysRevD.109.114019",
    journal = "Phys. Rev. D",
    volume = "109",
    number = "11",
    pages = "114019",
    year = "2024"
}

@article{Mutuk:2023oyz,
    author = "Mutuk, Halil",
    title = "{Masses and magnetic moments of doubly heavy tetraquarks via diffusion Monte Carlo method}",
    eprint = "2312.13383",
    archivePrefix = "arXiv",
    primaryClass = "hep-ph",
    doi = "10.1140/epjc/s10052-024-12736-3",
    journal = "Eur. Phys. J. C",
    volume = "84",
    number = "4",
    pages = "395",
    year = "2024"
}

@article{Wang:2023bek,
    author = "Wang, Fu-Lai and Luo, Si-Qiang and Liu, Xiang",
    title = "{Radiative decays and magnetic moments of the predicted Bc-like molecules}",
    eprint = "2303.04542",
    archivePrefix = "arXiv",
    primaryClass = "hep-ph",
    doi = "10.1103/PhysRevD.107.114017",
    journal = "Phys. Rev. D",
    volume = "107",
    number = "11",
    pages = "114017",
    year = "2023"
}

@article{Ozdem:2023rkx,
    author = {{\"O}zdem, Ula{\c{s}}},
    title = "{Analysis of the $Z_b(10650)$ state based on electromagnetic properties}",
    eprint = "2311.11327",
    archivePrefix = "arXiv",
    primaryClass = "hep-ph",
    doi = "10.1140/epjc/s10052-024-12408-2",
    journal = "Eur. Phys. J. C",
    volume = "84",
    number = "1",
    pages = "45",
    year = "2024"
}

@article{Ozdem:2023frj,
    author = {{\"O}zdem, Ulas},
    title = "{Magnetic and quadrupole moments of the , , and states in the diquark-antidiquark picture}",
    eprint = "2307.05028",
    archivePrefix = "arXiv",
    primaryClass = "hep-ph",
    doi = "10.1088/1674-1137/ad0110",
    journal = "Chin. Phys. C",
    volume = "48",
    number = "1",
    pages = "013101",
    year = "2024"
}

@article{Lei:2023ttd,
    author = "Lei, Ya-Ding and Li, Hao-Song",
    title = "{Electromagnetic properties of the Tcc+ molecular states}",
    eprint = "2312.01332",
    archivePrefix = "arXiv",
    primaryClass = "hep-ph",
    doi = "10.1103/PhysRevD.109.076014",
    journal = "Phys. Rev. D",
    volume = "109",
    number = "7",
    pages = "076014",
    year = "2024"
}

@article{Zhang:2021yul,
    author = "Zhang, Wen-Xuan and Xu, Hao and Jia, Duojie",
    title = "{Masses and magnetic moments of hadrons with one and two open heavy quarks: Heavy baryons and tetraquarks}",
    eprint = "2109.07040",
    archivePrefix = "arXiv",
    primaryClass = "hep-ph",
    doi = "10.1103/PhysRevD.104.114011",
    journal = "Phys. Rev. D",
    volume = "104",
    number = "11",
    pages = "114011",
    year = "2021"
}

@article{Azizi:2023gzv,
    author = {Azizi, K. and {\"O}zdem, U.},
    title = "{Exploring the magnetic dipole moments of $ {T}_{QQ\overline{q}\overline{s}} $ and $ {T}_{QQ\overline{s}\overline{s}} $ states in the framework of QCD light-cone sum rules}",
    eprint = "2301.07713",
    archivePrefix = "arXiv",
    primaryClass = "hep-ph",
    doi = "10.1007/JHEP03(2023)166",
    journal = "JHEP",
    volume = "03",
    pages = "166",
    year = "2023"
}

@article{Ozdem:2022eds,
    author = {{\"O}zdem, Ula{\c{s}}},
    title = "{Electromagnetic form factors of the Bc-like tetraquarks: Molecular and diquark-antidiquark pictures}",
    eprint = "2211.10169",
    archivePrefix = "arXiv",
    primaryClass = "hep-ph",
    doi = "10.1016/j.physletb.2023.137750",
    journal = "Phys. Lett. B",
    volume = "838",
    pages = "137750",
    year = "2023"
}

@article{Ozdem:2022kck,
    author = {{\"O}zdem, Ula{\c{s}}},
    title = "{Magnetic moments of the vector hidden-charm tetraquark states}",
    eprint = "2206.05196",
    archivePrefix = "arXiv",
    primaryClass = "hep-ph",
    doi = "10.1103/PhysRevD.105.114030",
    journal = "Phys. Rev. D",
    volume = "105",
    number = "11",
    pages = "114030",
    year = "2022"
}

@article{Xu:2020qtg,
    author = "Xu, Yong-Jiang and Liu, Yong-Lu and Huang, Ming-Qiu",
    title = "{The magnetic moment of $Z_{c}(3900)$ as an axial-vector molecular state}",
    eprint = "2007.15214",
    archivePrefix = "arXiv",
    primaryClass = "hep-ph",
    doi = "10.1140/epjc/s10052-020-08515-5",
    journal = "Eur. Phys. J. C",
    volume = "80",
    number = "10",
    pages = "953",
    year = "2020"
}

@article{Wang:2017dce,
    author = "Wang, Zhi-Gang",
    title = "{The magnetic moment of the $Z_c(3900)$ as an axialvector tetraquark state with QCD sum rules}",
    eprint = "1712.05664",
    archivePrefix = "arXiv",
    primaryClass = "hep-ph",
    doi = "10.1140/epjc/s10052-018-5794-0",
    journal = "Eur. Phys. J. C",
    volume = "78",
    number = "4",
    pages = "297",
    year = "2018"
}

@article{Ozdem:2022yhi,
    author = {{\"O}zdem, Ula{\c{s}}},
    title = "{Magnetic dipole moments of states}",
    eprint = "2203.07759",
    archivePrefix = "arXiv",
    primaryClass = "hep-ph",
    doi = "10.1088/1674-1137/ac8653",
    journal = "Chin. Phys. C",
    volume = "46",
    number = "11",
    pages = "113106",
    year = "2022"
}

@article{Wang:2023vtx,
    author = "Wang, Yi-Heng and Wei, Jia and An, Chun-Sheng and Deng, Cheng-Rong",
    title = "{$Z_{cs}$(4000)$^{+}$ and $Z_{cs}$(4220)$^{+}$ in a Multiquark Color Flux-Tube Model}",
    doi = "10.1088/0256-307X/40/2/021201",
    journal = "Chin. Phys. Lett.",
    volume = "40",
    number = "2",
    pages = "021201",
    year = "2023"
}

@article{Ozdem:2021hmk,
    author = {{\"O}zdem, Ula{\c{s}}},
    title = "{Magnetic moments of the doubly charged axial-vector Tcc++ states}",
    eprint = "2112.10402",
    archivePrefix = "arXiv",
    primaryClass = "hep-ph",
    doi = "10.1103/PhysRevD.105.054019",
    journal = "Phys. Rev. D",
    volume = "105",
    number = "5",
    pages = "054019",
    year = "2022"
}

@article{Azizi:2021aib,
    author = {Azizi, K. and {\"O}zdem, U.},
    title = "{Magnetic dipole moments of the Tcc+ and ZV++ tetraquark states}",
    eprint = "2109.02390",
    archivePrefix = "arXiv",
    primaryClass = "hep-ph",
    doi = "10.1103/PhysRevD.104.114002",
    journal = "Phys. Rev. D",
    volume = "104",
    number = "11",
    pages = "114002",
    year = "2021"
}

@article{Ozdem:2021hka,
    author = {{\"O}zdem, Ula{\c{s}} and Y{\i}ld{\i}r{\i}m, Ay{\c{s}}e Karadeniz},
    title = "{Magnetic dipole moments of the Zc(4020)+, Zc(4200)+, Zcs(4000)+ and Zcs(4220)+ states in light-cone QCD}",
    eprint = "2104.13074",
    archivePrefix = "arXiv",
    primaryClass = "hep-ph",
    doi = "10.1103/PhysRevD.104.054017",
    journal = "Phys. Rev. D",
    volume = "104",
    number = "5",
    pages = "054017",
    year = "2021"
}

@article{Xu:2020evn,
    author = "Xu, Yong-Jiang and Liu, Yong-Lu and Cui, Chun-Yu and Huang, Ming-Qiu",
    title = "{$\bar D^{(*)}_s D^{(*)}$ molecular state with $J^P$= $1^+$}",
    eprint = "2011.14313",
    archivePrefix = "arXiv",
    primaryClass = "hep-ph",
    doi = "10.1103/PhysRevD.104.094028",
    journal = "Phys. Rev. D",
    volume = "104",
    number = "9",
    pages = "094028",
    year = "2021"
}

@article{Ozdem:2021yvo,
    author = {{\"O}zdem, U. and Azizi, K.},
    title = "{Magnetic dipole moment of the $Z_{cs}(3985)$ state: diquark{\textendash}antidiquark and molecular pictures}",
    eprint = "2102.09231",
    archivePrefix = "arXiv",
    primaryClass = "hep-ph",
    doi = "10.1140/epjp/s13360-021-01977-w",
    journal = "Eur. Phys. J. Plus",
    volume = "136",
    number = "9",
    pages = "968",
    year = "2021"
}

@article{Ozdem:2017exj,
    author = "Ozdem, U. and Azizi, K.",
    title = "{Magnetic dipole moment of $Z_b(10610)$ in light-cone QCD}",
    eprint = "1709.09714",
    archivePrefix = "arXiv",
    primaryClass = "hep-ph",
    doi = "10.1103/PhysRevD.97.014010",
    journal = "Phys. Rev. D",
    volume = "97",
    number = "1",
    pages = "014010",
    year = "2018"
}

@article{Ozdem:2017jqh,
    author = "Ozdem, U. and Azizi, K.",
    title = "{Magnetic and quadrupole moments of the $Z_c(3900)$}",
    eprint = "1707.09612",
    archivePrefix = "arXiv",
    primaryClass = "hep-ph",
    doi = "10.1103/PhysRevD.96.074030",
    journal = "Phys. Rev. D",
    volume = "96",
    number = "7",
    pages = "074030",
    year = "2017"
}

@article{Ozdem:2024rrg,
    author = {{\"O}zdem, Ula{\c{s}}},
    title = "{Study on the electromagnetic properties of the $[sc] [{\bar{q}} {\bar{b}}]$ and $[sc] [{\bar{s}} {\bar{b}}]$ states with $J^P = 1^+$}",
    eprint = "2405.11036",
    archivePrefix = "arXiv",
    primaryClass = "hep-ph",
    doi = "10.1140/epjp/s13360-025-06016-6",
    journal = "Eur. Phys. J. Plus",
    volume = "140",
    number = "2",
    pages = "105",
    year = "2025"
}

@article{Mutuk:2024vzv,
    author = "Mutuk, Halil",
    title = "{Doubly-charged Tcc++ states in the dynamical diquark model}",
    eprint = "2401.02788",
    archivePrefix = "arXiv",
    primaryClass = "hep-ph",
    doi = "10.1103/PhysRevD.110.034025",
    journal = "Phys. Rev. D",
    volume = "110",
    number = "3",
    pages = "034025",
    year = "2024"
}

@article{Wang:2016dzu,
    author = "Wang, Guang-Juan and Chen, Rui and Ma, Li and Liu, Xiang and Zhu, Shi-Lin",
    title = "{Magnetic moments of the hidden-charm pentaquark states}",
    eprint = "1605.01337",
    archivePrefix = "arXiv",
    primaryClass = "hep-ph",
    doi = "10.1103/PhysRevD.94.094018",
    journal = "Phys. Rev. D",
    volume = "94",
    number = "9",
    pages = "094018",
    year = "2016"
}

@article{Ortiz-Pacheco:2018ccl,
    author = "Ortiz-Pacheco, Emmanuel and Bijker, Roelof and Fern{\'a}ndez-Ram{\'\i}rez, C{\'e}sar",
    title = "{Hidden charm pentaquarks: mass spectrum, magnetic moments, and photocouplings}",
    eprint = "1808.10512",
    archivePrefix = "arXiv",
    primaryClass = "nucl-th",
    doi = "10.1088/1361-6471/ab096d",
    journal = "J. Phys. G",
    volume = "46",
    number = "6",
    pages = "065104",
    year = "2019"
}

@article{Xu:2020flp,
    author = "Xu, Yong-Jiang and Liu, Yong-Lu and Huang, Ming-Qiu",
    title = "{The magnetic moment of $P_{c}(4312)$ as a $\bar{D}\Sigma _{c}$ molecular state}",
    eprint = "2008.07937",
    archivePrefix = "arXiv",
    primaryClass = "hep-ph",
    doi = "10.1140/epjc/s10052-021-09211-8",
    journal = "Eur. Phys. J. C",
    volume = "81",
    number = "5",
    pages = "421",
    year = "2021"
}

@article{Li:2021ryu,
    author = "Li, Ming-Wei and Liu, Zhan-Wei and Sun, Zhi-Feng and Chen, Rui",
    title = "{Magnetic moments and transition magnetic moments of Pc and Pcs states}",
    eprint = "2106.15053",
    archivePrefix = "arXiv",
    primaryClass = "hep-ph",
    doi = "10.1103/PhysRevD.104.054016",
    journal = "Phys. Rev. D",
    volume = "104",
    number = "5",
    pages = "054016",
    year = "2021"
}

@article{Ozdem:2023htj,
    author = {{\"O}zdem, Ula{\c{s}}},
    title = "{Electromagnetic properties of D{\textasciimacron}({\textasteriskcentered}){\ensuremath{\Xi}}c', D{\textasciimacron}({\textasteriskcentered}){\ensuremath{\Lambda}}c, D{\textasciimacron}s({\textasteriskcentered}){\ensuremath{\Lambda}}c and D{\textasciimacron}s({\textasteriskcentered}){\ensuremath{\Xi}}c pentaquarks}",
    eprint = "2303.10649",
    archivePrefix = "arXiv",
    primaryClass = "hep-ph",
    doi = "10.1016/j.physletb.2023.138267",
    journal = "Phys. Lett. B",
    volume = "846",
    pages = "138267",
    year = "2023"
}

@article{Wang:2023iox,
    author = "Wang, Fu-Lai and Liu, Xiang",
    title = "{Higher molecular P{\ensuremath{\psi}}s{\ensuremath{\Lambda}}/{\ensuremath{\Sigma}} pentaquarks arising from the {\ensuremath{\Xi}}c(',*)D{\textasciimacron}1/{\ensuremath{\Xi}}c(',*)D{\textasciimacron}2* interactions}",
    eprint = "2307.08276",
    archivePrefix = "arXiv",
    primaryClass = "hep-ph",
    doi = "10.1103/PhysRevD.108.054028",
    journal = "Phys. Rev. D",
    volume = "108",
    number = "5",
    pages = "054028",
    year = "2023"
}

@article{Ozdem:2022kei,
    author = {{\"O}zdem, Ula{\c{s}}},
    title = "{Investigation of magnetic moment of Pcs(4338) and Pcs(4459) pentaquark states}",
    eprint = "2208.07684",
    archivePrefix = "arXiv",
    primaryClass = "hep-ph",
    doi = "10.1016/j.physletb.2022.137635",
    journal = "Phys. Lett. B",
    volume = "836",
    pages = "137635",
    year = "2023"
}

@article{Gao:2021hmv,
    author = "Gao, Feng and Li, Hao-Song",
    title = "{Magnetic moments of hidden-charm strange pentaquark states*}",
    eprint = "2112.01823",
    archivePrefix = "arXiv",
    primaryClass = "hep-ph",
    doi = "10.1088/1674-1137/ac8651",
    journal = "Chin. Phys. C",
    volume = "46",
    number = "12",
    pages = "123111",
    year = "2022"
}

@article{Ozdem:2024rch,
    author = {{\"O}zdem, Ula{\c{s}}},
    title = "{Shedding light on the nature of the $P_{cs}(4459)$ pentaquark state}",
    eprint = "2411.11442",
    archivePrefix = "arXiv",
    primaryClass = "hep-ph",
    doi = "10.1103/PhysRevD.111.074038",
    journal = "Phys. Rev. D",
    volume = "111",
    number = "7",
    pages = "074038",
    year = "2025"
}

@article{Guo:2023fih,
    author = "Guo, Fei and Li, Hao-Song",
    title = "{Analysis of the hidden-charm pentaquark states based on magnetic moment and transition magnetic moment}",
    eprint = "2304.10981",
    archivePrefix = "arXiv",
    primaryClass = "hep-ph",
    doi = "10.1140/epjc/s10052-024-12699-5",
    journal = "Eur. Phys. J. C",
    volume = "84",
    number = "4",
    pages = "392",
    year = "2024"
}

@article{Ozdem:2022iqk,
    author = {{\"O}zdem, Ula{\c{s}}},
    title = "{Magnetic moments of pentaquark states in light-cone sum rules}",
    doi = "10.1140/epja/s10050-022-00700-2",
    journal = "Eur. Phys. J. A",
    volume = "58",
    number = "3",
    pages = "46",
    year = "2022"
}

@article{Wang:2022nqs,
    author = "Wang, Fu-Lai and Luo, Si-Qiang and Zhou, Hong-Yan and Liu, Zhan-Wei and Liu, Xiang",
    title = "{Exploring the electromagnetic properties of the {\ensuremath{\Xi}}c(',*)D{\textasciimacron}s* and {\ensuremath{\Omega}}c(*)D{\textasciimacron}s* molecular states}",
    eprint = "2210.02809",
    archivePrefix = "arXiv",
    primaryClass = "hep-ph",
    doi = "10.1103/PhysRevD.108.034006",
    journal = "Phys. Rev. D",
    volume = "108",
    number = "3",
    pages = "034006",
    year = "2023"
}

@article{Wang:2022tib,
    author = "Wang, Fu-Lai and Zhou, Hong-Yan and Liu, Zhan-Wei and Liu, Xiang",
    title = "{What can we learn from the electromagnetic properties of hidden-charm molecular pentaquarks with single strangeness?}",
    eprint = "2208.10756",
    archivePrefix = "arXiv",
    primaryClass = "hep-ph",
    doi = "10.1103/PhysRevD.106.054020",
    journal = "Phys. Rev. D",
    volume = "106",
    number = "5",
    pages = "054020",
    year = "2022"
}

@article{Ozdem:2024jty,
    author = {{\"O}zdem, Ula{\c{s}}},
    title = "{Analysis of the isospin eigenstate $\bar{D} \Sigma _c$, $\bar{D}^{*} \Sigma _c$, and $\bar{D} \Sigma _c^{*}$ pentaquarks by their electromagnetic properties}",
    eprint = "2401.12678",
    archivePrefix = "arXiv",
    primaryClass = "hep-ph",
    doi = "10.1140/epjc/s10052-024-13124-7",
    journal = "Eur. Phys. J. C",
    volume = "84",
    number = "8",
    pages = "769",
    year = "2024"
}

@article{Li:2024wxr,
    author = "Li, Hao-Song and Guo, Fei and Lei, Ya-Ding and Gao, Feng",
    title = "{Magnetic moments and axial charges of the octet hidden-charm molecular pentaquark family}",
    eprint = "2401.14767",
    archivePrefix = "arXiv",
    primaryClass = "hep-ph",
    doi = "10.1103/PhysRevD.109.094027",
    journal = "Phys. Rev. D",
    volume = "109",
    number = "9",
    pages = "094027",
    year = "2024"
}

@article{Li:2024jlq,
    author = "Li, Hao-Song",
    title = "{Molecular pentaquark magnetic moments in heavy pentaquark chiral perturbation theory}",
    eprint = "2401.14759",
    archivePrefix = "arXiv",
    primaryClass = "hep-ph",
    doi = "10.1103/PhysRevD.109.114039",
    journal = "Phys. Rev. D",
    volume = "109",
    number = "11",
    pages = "114039",
    year = "2024"
}

@article{Ozdem:2024yel,
    author = {{\"O}zdem, Ula{\c{s}}},
    title = "{Investigation on the electromagnetic properties of the $ D^{(*)} \Sigma _c^{(*)}$ molecules}",
    eprint = "2405.07273",
    archivePrefix = "arXiv",
    primaryClass = "hep-ph",
    doi = "10.1140/epja/s10050-024-01477-2",
    journal = "Eur. Phys. J. A",
    volume = "61",
    number = "1",
    pages = "10",
    year = "2025"
}

@article{Ozdem:2024rqx,
    author = {{\"O}zdem, Ula{\c{s}}},
    title = "{Elucidating the nature of hidden-charm pentaquark states with spin-32 through their electromagnetic form factors}",
    eprint = "2402.03802",
    archivePrefix = "arXiv",
    primaryClass = "hep-ph",
    doi = "10.1016/j.physletb.2024.138551",
    journal = "Phys. Lett. B",
    volume = "851",
    pages = "138551",
    year = "2024"
}

@article{Mutuk:2024ltc,
    author = "Mutuk, Halil and Kang, Xian-Wei",
    title = "{Unveiling the structure of hidden-bottom strange pentaquarks via magnetic moments}",
    eprint = "2405.07066",
    archivePrefix = "arXiv",
    primaryClass = "hep-ph",
    doi = "10.1016/j.physletb.2024.138772",
    journal = "Phys. Lett. B",
    volume = "855",
    pages = "138772",
    year = "2024"
}

@article{Mutuk:2024jxf,
    author = "Mutuk, Halil",
    title = "{Magnetic moments of hidden-bottom pentaquark states}",
    eprint = "2403.16616",
    archivePrefix = "arXiv",
    primaryClass = "hep-ph",
    doi = "10.1140/epjc/s10052-024-13263-x",
    journal = "Eur. Phys. J. C",
    volume = "84",
    number = "8",
    pages = "874",
    year = "2024"
}

@article{Zhu:2025abk,
    author = "Zhu, Sheng-He and Wang, Fu-Lai and Liu, Xiang",
    title = "{Electromagnetic characteristics as probes into the inner structures of the predicted $\Xi_c^{(',*)}D^{(*)}_s$ molecular states}",
    eprint = "2510.18492",
    archivePrefix = "arXiv",
    primaryClass = "hep-ph",
    month = "10",
    year = "2025"
}

@article{Ozdem:2025ncd,
    author = {{\"O}zdem, Ula{\c{s}}},
    title = "{Unveiling the electromagnetic structure and intrinsic dynamics of spin-3/2 hidden-charm pentaquarks: A comprehensive QCD analysis}",
    eprint = "2504.13488",
    archivePrefix = "arXiv",
    primaryClass = "hep-ph",
    doi = "10.1088/1674-1137/ade95a",
    journal = "Chin. Phys.",
    volume = "49",
    number = "10",
    pages = "103106",
    year = "2025"
}

@article{Ozdem:2025ion,
    author = {{\"O}zdem, U.},
    title = "{Electromagnetic form factors: A window into the $D\Lambda_c$, $D^*\Lambda_c$, and $D\Lambda_c^*$ molecular structure}",
    eprint = "2511.16052",
    archivePrefix = "arXiv",
    primaryClass = "hep-ph",
    month = "11",
    year = "2025"
}

@article{Mutuk:2024ach,
    author = "Mutuk, Halil",
    title = "{Magnetic moments of hidden-charm pentaquarks in the diquark{\textendash}diquark{\textendash}antiquark scheme}",
    eprint = "2411.16486",
    archivePrefix = "arXiv",
    primaryClass = "hep-ph",
    doi = "10.1016/j.cjph.2025.07.030",
    journal = "Chin. J. Phys.",
    volume = "97",
    pages = "1406--1414",
    year = "2025"
}

@article{Ozdem:2025jda,
    author = {{\"O}zdem, Ula{\c{s}}},
    title = "{Electromagnetic tomography of spin-$\frac{3}{2}$ hidden-charm strange pentaquarks}",
    eprint = "2510.26893",
    archivePrefix = "arXiv",
    primaryClass = "hep-ph",
    month = "10",
    year = "2025"
}

@article{Brodsky:1992px,
    author = "Brodsky, Stanley J. and Hiller, John R.",
    title = "{Universal properties of the electromagnetic interactions of spin one systems}",
    reportNumber = "SLAC-PUB-5763",
    doi = "10.1103/PhysRevD.46.2141",
    journal = "Phys. Rev. D",
    volume = "46",
    pages = "2141--2149",
    year = "1992"
}

@article{Wang:2010sh,
    author = "Wang, Zhi-Gang",
    title = "{Analysis of the scalar and axial-vector heavy diquark states with QCD sum rules}",
    eprint = "1008.4449",
    archivePrefix = "arXiv",
    primaryClass = "hep-ph",
    doi = "10.1140/epjc/s10052-010-1524-y",
    journal = "Eur. Phys. J. C",
    volume = "71",
    pages = "1524",
    year = "2011"
}

@article{Kleiv:2013dta,
    author = "Kleiv, R. T. and Steele, T. G. and Zhang, Ailin and Blokland, Ian",
    title = "{Heavy-light diquark masses from QCD sum rules and constituent diquark models of tetraquarks}",
    eprint = "1304.7816",
    archivePrefix = "arXiv",
    primaryClass = "hep-ph",
    doi = "10.1103/PhysRevD.87.125018",
    journal = "Phys. Rev. D",
    volume = "87",
    number = "12",
    pages = "125018",
    year = "2013"
}

@article{Balitsky:1987bk,
    author = "Balitsky, I. I. and Braun, Vladimir M.",
    title = "{Evolution Equations for QCD String Operators}",
    reportNumber = "LENINGRAD-87-1351",
    doi = "10.1016/0550-3213(89)90168-5",
    journal = "Nucl. Phys. B",
    volume = "311",
    pages = "541--584",
    year = "1989"
}

@article{Belyaev:1985wza,
    author = "Belyaev, V. M. and Blok, B. Yu.",
    title = "{CHARMED BARYONS IN QUANTUM CHROMODYNAMICS}",
    doi = "10.1007/BF01560689",
    journal = "Z. Phys. C",
    volume = "30",
    pages = "151",
    year = "1986"
}

@article{Li:2020rcg,
    author = {Li, Hua-Dong and L{\"u}, Cai-Dian and Wang, Chao and Wang, Yu-Ming and Wei, Yan-Bing},
    title = "{QCD calculations of radiative heavy meson decays with subleading power corrections}",
    eprint = "2002.03825",
    archivePrefix = "arXiv",
    primaryClass = "hep-ph",
    doi = "10.1007/JHEP04(2020)023",
    journal = "JHEP",
    volume = "04",
    pages = "023",
    year = "2020"
}

@article{Ball:2002ps,
    author = "Ball, Patricia and Braun, V. M. and Kivel, N.",
    title = "{Photon distribution amplitudes in QCD}",
    eprint = "hep-ph/0207307",
    archivePrefix = "arXiv",
    reportNumber = "IPPP-02-40, DCPT-02-80",
    doi = "10.1016/S0550-3213(02)01017-9",
    journal = "Nucl. Phys. B",
    volume = "649",
    pages = "263--296",
    year = "2003"
}

@article{Ozdem:2022vip,
    author = {{\"O}zdem, Ulas},
    title = "{Electromagnetic properties of doubly heavy pentaquark states}",
    eprint = "2201.00979",
    archivePrefix = "arXiv",
    primaryClass = "hep-ph",
    doi = "10.1140/epjp/s13360-022-03125-4",
    journal = "Eur. Phys. J. Plus",
    volume = "137",
    pages = "936",
    year = "2022"
}

@article{ParticleDataGroup:2024cfk,
    author = "Navas, S. and others",
    collaboration = "Particle Data Group",
    title = "{Review of particle physics}",
    doi = "10.1103/PhysRevD.110.030001",
    journal = "Phys. Rev. D",
    volume = "110",
    number = "3",
    pages = "030001",
    year = "2024"
}

@article{Ioffe:2005ym,
    author = "Ioffe, B. L.",
    title = "{QCD at low energies}",
    eprint = "hep-ph/0502148",
    archivePrefix = "arXiv",
    doi = "10.1016/j.ppnp.2005.05.001",
    journal = "Prog. Part. Nucl. Phys.",
    volume = "56",
    pages = "232--277",
    year = "2006"
}

@article{Narison:2018nbv,
    author = "Narison, Stephan",
    editor = "Narison, St{\'e}phan",
    title = "{$\overline{\rm m}_{c,b,}<\alpha_sG^2>$ and $\alpha_s$ from Heavy Quarkonia}",
    doi = "10.1016/j.nuclphysbps.2018.12.026",
    journal = "Nucl. Part. Phys. Proc.",
    volume = "300-302",
    pages = "153--164",
    year = "2018"
}
\bibliographystyle{elsarticle-num}

\end{document}